\documentstyle{mn}

\newif\ifAMStwofonts
\AMStwofontstrue

\newcommand{\ton}{TON~S~180}
\newcommand{\pks}{PKS~0558--504}
\newcommand{\ark}{Ark~564}
\newcommand{\ngc}{NGC~4051}
\newcommand{\mrk}{Mrk~335}
\newcommand{\pg}{PG~1244+026}
\newcommand{\kev}{keV}
\newcommand{\ergcms}{erg~cm~s$^{-1}$}

\newcommand{\fe}{Fe~K$\alpha$}
\newcommand{\etal}{et al.}

\ifoldfss
  \ifCUPmtlplainloaded \else
    \NewTextAlphabet{textbfit} {cmbxti10} {}
    \NewTextAlphabet{textbfss} {cmssbx10} {}
    \NewMathAlphabet{mathbfit} {cmbxti10} {} 
    \NewMathAlphabet{mathbfss} {cmssbx10} {} 
  \fi
  \ifAMStwofonts
    \ifCUPmtlplainloaded \else
      \NewSymbolFont{upmath} {eurm10}
      \NewSymbolFont{AMSa} {msam10}
      \NewMathSymbol{\upi}     {0}{upmath}{19}
      \NewMathSymbol{\umu}     {0}{upmath}{16}
      \NewMathSymbol{\upartial}{0}{upmath}{40}
      \NewMathSymbol{\leqslant}{3}{AMSa}{36}
      \NewMathSymbol{\geqslant}{3}{AMSa}{3E}

      \let\leq=\leqslant 
       
    \fi
  \fi
\fi 

\ifnfssone
  \newmathalphabet{\mathit}
  \addtoversion{normal}{\mathit}{cmr}{m}{it}
  \addtoversion{bold}{\mathit}{cmr}{bx}{it}
  \newmathalphabet{\mathbfit} 
  \addtoversion{normal}{\mathbfit}{cmr}{bx}{it}
  \addtoversion{bold}{\mathbfit}{cmr}{bx}{it}
  \newmathalphabet{\mathbfss} 
  \addtoversion{normal}{\mathbfss}{cmss}{bx}{n}
  \addtoversion{bold}{\mathbfss}{cmss}{bx}{n}
  \ifAMStwofonts
    \ifCUPmtlplainloaded \else
      \UseAMStwoboldmath
      \makeatletter
      \new@mathgroup\upmath@group
      \define@mathgroup\mv@normal\upmath@group{eur}{m}{n}
      \define@mathgroup\mv@bold\upmath@group{eur}{b}{n}
      \edef\UPM{\hexnumber\upmath@group}
      \new@mathgroup\amsa@group
      \define@mathgroup\mv@normal\amsa@group{msa}{m}{n}
      \define@mathgroup\mv@bold\amsa@group{msa}{m}{n}
      \edef\AMSa{\hexnumber\amsa@group}
      \makeatother
      \mathchardef\upi="0\UPM19
      \mathchardef\umu="0\UPM16
      \mathchardef\upartial="0\UPM40
      \mathchardef\leqslant="3\AMSa36
      \mathchardef\geqslant="3\AMSa3E

      \let\leq=\leqslant 

    \fi
  \fi
\fi 

\ifnfsstwo
  \DeclareMathAlphabet{\mathbfit}{OT1}{cmr}{bx}{it}
  \SetMathAlphabet\mathbfit{bold}{OT1}{cmr}{bx}{it}
  \DeclareMathAlphabet{\mathbfss}{OT1}{cmss}{bx}{n}
  \SetMathAlphabet\mathbfss{bold}{OT1}{cmss}{bx}{n}
  \ifAMStwofonts
    \ifCUPmtlplainloaded \else
      \DeclareSymbolFont{UPM}{U}{eur}{m}{n}
      \SetSymbolFont{UPM}{bold}{U}{eur}{b}{n}
      \DeclareSymbolFont{AMSa}{U}{msa}{m}{n}
      \DeclareMathSymbol{\upi}{0}{UPM}{"19}
      \DeclareMathSymbol{\umu}{0}{UPM}{"16}
      \DeclareMathSymbol{\upartial}{0}{UPM}{"40}
      \DeclareMathSymbol{\leqslant}{3}{AMSa}{"36}
      \DeclareMathSymbol{\geqslant}{3}{AMSa}{"3E}

      \let\leq=\leqslant 

    \fi
  \fi
\fi 

\ifCUPmtlplainloaded \else
  \ifAMStwofonts \else 
    \def\upi{\pi}
    \def\umu{\mu}
    \def\upartial{\partial}
  \fi
\fi

\title[Ionised discs in 5 NLS1]
  {Evidence for ionised accretion discs in five narrow-line Seyfert 1 galaxies}
\author[D.\ R.\ Ballantyne, K.\ Iwasawa \& A.\ C.\ Fabian]
  {D.~R.~Ballantyne\thanks{drb@ast.cam.ac.uk}, 
  K.~Iwasawa \& A.~C.~Fabian\\
  Institute of Astronomy, Madingley Road, Cambridge CB3 0HA UK}
\date{2000 June 1}
\pagerange{\pageref{firstpage}--\pageref{lastpage}}
\pubyear{2000}

\input{psfig.sty}

\begin{document}

\label{firstpage}

\maketitle

\begin{abstract}
We present the results of fitting \textit{ASCA} spectra of six
narrow-line Seyfert 1 (NLS1) galaxies with the ionised reflection
models of Ross \& Fabian (1993). We find that five of the galaxies
(\ton, \pks, \ark, \mrk\ and \pg) are well fit by the ionised disc
model, and these are often better fits than the alternative models
considered. The sixth galaxy, \ngc, has additional spectral complexity
that cannot be well described by a simple ionised disc model or any of
the other alternative models. The highest luminosity NLS1 considered,
\pks, does not have a well constrained ionisation parameter or
reflection fraction. This is because it is difficult to distinguish
between highly ionised, highly reflective discs and moderately ionised
discs with low reflection fractions. The four galaxies with well
constrained fit parameters are consistent with having inclination
angles lying between 15 and 30~degrees. Furthermore, these four
sources are consistent with having a disc emissivity law that varies
as $r^{-2.5}$. These last two properties are also typical of
broad-line Seyfert 1 galaxies. We find little or no indication of a
correlation between the reflection fraction and the photon index of
the underlying continuum. All six of the NLS1s we considered show
evidence for a broad \fe\ line, but none of the line centroids are
consistent with emission from highly ionised Fe. This is most likely
due to the line being redshifted because of relativistic effects. We
note that sources with larger ionisation parameters tend to have
larger \fe\ EWs. We interpret this as evidence that ionised Fe
features are making their presence felt in the spectra. Since most of
our sources have steep spectra, highly ionised features are predicted
even by the new variable density reflection models. The one source we
analyse with a photon index less than two has the lowest ionisation
parameter in our sample.  We conclude that our result is the strongest
evidence yet that NLS1 might have ionised accretion discs. This result
gives further weight to the hypothesis that these objects contain
rapidly accreting black holes.
 
\end{abstract}


\section{Introduction}
\label{sect:intro}
One of the most basic difficulties in attempting to understand the physics
of Active Galactic Nuclei (AGN), is to account for the wide range of
properties (i.e., broad lines, radio properties, etc.) that are observed 
over the whole menagerie of AGN classes. Although it is quite
likely that we are observing basically the same phenomenon through
different observing angles (e.g., Antonucci 1993), not all AGN can be
unified through this picture.  Certain classes of AGN seem to be
extreme examples of the phenomenon, and cannot fit into the simple
unified scheme. It is hoped that such objects would be dominated by
one physical property in their central engines, and would therefore be
simpler to model than a ``typical'' AGN.  An example of this type of
extreme AGN is the Narrow Line Seyfert 1 (NLS1) class of galaxies.

NLS1 galaxies, like most AGN, are categorized by their optical
emission properties. A NLS1 galaxy has the following characteristics: FWHM
H$\beta < 2000$~km~s$^{-1}$, [O III]/H$\beta < 3$, and strong Fe II
emission (Osterbrock \& Pogge 1985). These properties mean that a NLS1
has narrower permitted lines (i.e., broad lines) then a typical
Seyfert 1 galaxy. These galaxies also exhibit unusual X-ray properties
(Brandt 1999; Leighly 1999a; Leighly 1999b; Vaughan \etal\ 1999b). In
the soft X-ray band (usually taken to mean the {\it ROSAT} band: 0.1--2.4~keV) 
these galaxies show very steep spectra due to a significant soft excess, 
with a power-law photon index ($\Gamma$; where the photon flux
$\propto E^{-\Gamma}$) sometimes exceeding 3 (Boller, Brandt \& Fink
1996).  In the hard X-ray band, the steep power-law flattens out to a
typical $\Gamma \approx$ 2.1--2.4 (Brandt, Mathur \& Elvis 1997;
Leighly 1999b; Vaughan \etal\ 1999b), which is significantly steeper
than the spectra of broad-line Seyfert 1 objects.  The
soft excess is well fit by a model that is suggestive of thermal
emission from a disc. NLS1 also exhibit more rapid and extreme
variability in the X-ray bands than their broad-line counterparts
(Turner \etal\ 1999; Leighly 1999a).

The model that is currently the most successful in explaining these
properties is one which proposes that NLS1 galaxies have smaller black
hole masses than typical Seyfert galaxies (e.g., Boller \etal\
1996; Laor \etal\ 1997). This would explain the optical line
properties, because the material in the broad line region would have 
smaller velocities and so their lines will have a smaller Doppler
width.  However, since these galaxies have about the same luminosity
as broad-line Seyfert galaxies, then they must be emitting a higher
fraction of their Eddington luminosity. This implies a higher
accretion rate and a hotter accretion disc, which would cause the
thermal emission of the disc to shift into the soft X-ray band. The
steeper high energy $\Gamma$ could then be explained by Compton
cooling of the hard X-ray emitting corona by the soft emission from
the disc. The extreme variability would just be a result of the
smaller black hole mass because the primary emission region would be
smaller; however, it is difficult to rule out the possibility of a beamed 
component. An analogy between NLS1 and Galactic black hole candidates
emitting in their soft/high state has often been stated (e.g., Pounds,
Done \& Osborne 1995), although its usefulness has been debated.

Hard X-ray observations of NLS1 can test this model. Models of \fe\
line emission at various accretion rates by Matt, Fabian \& Ross
(1993) have shown that at modest accretion rates, the accretion disc
will have an ionised skin on top. This result assumes that the
illuminating X-ray flux depends on the accretion rate, which is
reasonable if the hard X-ray emitting corona does result from a
process like magnetic flaring (e.g., Galeev, Rosner \& Vaiana 1979;
Haardt, Maraschi \& Ghisellini 1994; Svensson 1996).  Preliminary evidence 
for ionised accretion discs based on
\textit{ASCA} data has been found in two NLS1 (TON~S~180: Comastri
\etal\ (1998) and Turner, George \& Nandra (1998); Ark~564: Vaughan \etal\
(1999a), but see Turner, George \& Netzer (1999) for an alternative 
interpretation). However, these
conclusions were made either by estimating the energy centroid of the
observed \fe\ line or Fe K edge and ignoring the rest of the continuum, or 
by using the
\textsc{pexriv} ionised disc model of Magdziarz \& Zdziarski
(1995). Ionised reflection has observable effects over the whole
continuum, so better constraints on the ionisation state can be made by
fitting ionised disc models over the entire observed energy
range. However, as shown by Ross, Fabian \& Young (1999), the
\textsc{pexriv} model is inaccurate as the disc becomes highly
ionised. A further limitation of this model is that the \fe\ line has
to be added in separately.

This paper presents the results of fitting the ionised disc models of
Ross \& Fabian (1993; hereafter RF) to \textit{ASCA} observations of 6
NLS1 galaxies.  These models not only include the \fe\ line, but also
spectral features (emission lines and recombination continua) at lower
energies. We take advantage of this information by fitting the
models over a wide range of energy. The data selection and reduction
procedure is described in Section~\ref{sect:data}. We briefly outline
the properties of our models in Section~\ref{sect:model}, and then
move on to report the results of our fits in
Section~\ref{sect:res}. Finally, we discuss our results in 
Section~\ref{sect:discuss}, and summarise our conclusions in 
Section~\ref{sect:concl}.

\section{Observations and data reduction}
\label{sect:data}

A summary of \textit{ASCA} observations of the sample galaxies, as
well as other relevant data, is shown in Table~\ref{table:prop}. The
spectral data from the four detectors were reduced from the event
files taken from the \textit{ASCA} archive maintained by the
\textit{ASCA} Guest Observer Facility (\textit{ASCA} GOF) at
NASA/Goddard Space Flight Center, using \textsc{ftools} 4.2 and the
standard calibration. Source photons were collected from a region with
a radius of typically 4 arcmin for the Solid state Imaging
Spectrometer (SIS; S0 and S1) and 5 arcmin for the Gas Imaging
Spectrometer (GIS; G2 and G3), while background data were taken from a
source-free region on the detectors in the same observations.  All the
targets show significant X-ray variability during the observations.
The \textit{ASCA} light curves of our sample galaxies have been 
published elsewhere (Leighly 1999a). We present a
spectral analysis of time-averaged data of individual sources
integrated over each observation.

The SIS was operating with 2CCD chips switching between Faint and
Bright modes for the observations of \mrk\ and \pg. The Faint mode
data of these observations have been converted to the Bright mode
format to add together with the original Bright mode data. For the
others, the SIS was operating with 1CCD Faint mode throughout. These
Faint mode data have been converted to a format called ``Bright2'' for
analysis.

Response matrices for the SIS were generated by \textsc{sisrmg}
version 1.1.  Version 4.0 matrices provided by the GIS team were used
for the GIS.  The effective areas of the source spectra were computed
with \textsc{ascaarf} version 2.73. For observations carried out after
1994, the efficiency of the SIS, typically in the energy range below
1~keV, has been noticed to be decreasing continuously. Since the 
correction for this effect has not been
implemented in \textsc{ascaarf} at this stage, the effective area of
the SIS in the low energy range is likely to be overestimated for some
targets in our sample. We therefore restrict our spectral analysis to
the energy range above 1~keV, except the data for \mrk\ and \ngc\
which were observed when the SIS degradation was insignificant.

\begin{table*}
\begin{minipage}{170mm}
\caption{Summary of \textit{ASCA} observations and other relevant properties 
of the sample galaxies. $N_H$ is the Galactic absorbing column in units of 
10$^{20}$~cm$^{-2}$, and $L_{2-10}$ is the 2--10~\kev\ luminosity in units 
of 10$^{43}$~ergs~s$^{-1}$ (assuming $H_0$=50~km~s$^{-1}$~Mpc$^{-1}$).}
\label{table:prop}
\begin{tabular}{@{}lccccccccc}
Galaxy & $z$ & Date & SIS mode & Exposure$^1$ & Count rate (1--10 keV)$^2$ & $N_H^{a}$ & $L_{2-10}^{a}$ & {\em ROSAT} $\Gamma^{a}$ \\
 & & & & SIS/GIS (ks) & SIS/GIS (cps) \\ \hline
\mrk & 0.025 & 1993 Dec 9 & 2CCD Faint/Bright & 25.1/20.4 & 0.475/0.294 & 1.50 & 8.9 & 3.04 \\
\ngc & 0.0024 & 1994 Jun 7--9 & 1CCD Faint & 81.7/71.1 & 0.854/0.472 & 1.31 & 0.056 & 2.84 \\
\pg & 0.048 & 1996 Jul 1--3 & 2CCD Faint/Bright & 49.6/38.8 & 0.165/0.085 & 1.93 & 2.62 & 3.26 \\
\ton & 0.062 & 1996 Jul 10--11 & 1CCD Faint & 54.2/49.2 & 0.306/0.158 & 1.50 & 8.9 & 3.04 \\
\ark & 0.024 & 1996 Dec 23--24 & 1CCD Faint & 54.8/50.8 & 1.329/0.657 & 6.40 & 5.5 & 3.47 \\
\pks & 0.137 & 1996 Sep 5--6 & 1CCD Faint & 44.3/34.4 & 0.660/0.359 & 4.39 & 124 & 2.89 \\
\end{tabular}

\medskip
$^1$ Good exposure time of single detector of each type.

$^2$ Mean count rate in the 1--10~keV band obtained from the
S0 (for the SIS) and G2 (for the GIS) detectors.

$^a$ data taken from Leighly (1999b).

\end{minipage}
\end{table*}

\section{Ionised Disc Model}
\label{sect:model}
We employed the ionised disc models described in detail by RF and Ross
\etal\ (1999), and so we will only give a brief outline here. A slab
of gas consisting of solar Morrison \& McCammon (1983) abundances, and 
with a constant hydrogen number density of $n_H = 10^{15}$~cm$^{-3}$, is 
illuminated by an X-ray power-law with flux $F_x$ (defined over 
0.01--100~\kev) and photon index $\Gamma$. The computed reflection 
spectrum is then multiplied by a factor $R$, where $R$ is the reflected 
fraction, and then added to the illuminating spectrum to find the model 
observed spectrum.

The important quantity in determining the structure of the reflected
spectrum is the ionisation parameter,
\begin{equation}
\label{eq:ip}
\xi= \frac{4 \pi F_x}{n_H}.
\end{equation}
The larger $\xi$ is, the more ionised the gas is, and this will affect
the strength and width of the features in the reflected spectrum like
the \fe\ line and the various absorption edges (Matt \etal\ 1993, 1996).

We computed a three-dimensional grid of models with $0.0 \leq R \leq
2.0$, $1.7 \leq \Gamma \leq 3.0$, and $1.0 \leq \log \xi \leq
6.0$. The ionisation parameter was varied by changing only the
incident flux. We did not include the influence of soft radiation from
the disc itself, but since we are mainly dealing with energies greater
than 1~\kev\ (Sections~\ref{sect:data} \&~\ref{sub:res-fit}), this should 
not be an important effect. 

There are some important deficiencies in the RF models. Foremost among
them is the assumption of constant density in the reflecting slab.
Recently, Nayakshin, Kazanas \& Kallman (2000) presented ionised
reflection models where hydrostatic equilibrium is solved along with
the ionisation and radiation structure. They find that, due to a
thermal instability, only a very thin layer (with Thomson optical depth, 
$\tau_t \la 1$) at the top of the slab is
highly ionised, and the reflection spectrum is dominated by the cool,
neutral material underneath. If this is the case, then it would be
inappropriate to fit data with the reflection spectra predicted by
constant density models, such as the RF model used here, because they
predict the presence of highly ionised features. However, all is not
lost, because the models of Nayakshin \etal\ (2000) show that if the disc is
radiation pressure dominated and the illuminating power-law has
$\Gamma \ga 2$ then ionised features can be seen in the reflection
spectrum. In that case, our fits to these NLS1 sources might be
reasonable because most of them have $\Gamma > 2$. Evidence for this
interpretation is presented in Section~\ref{sect:discuss}.

\section{Results}
\label{sect:res}
\subsection{Spectral fitting}
\label{sub:res-fit}
Data from all four detectors were used in the spectral fitting.
However, as mentioned in Section~\ref{sect:data}, the SIS detectors' 
efficiency below
1~\kev\ has slowly been degrading since late 1994. This is not
corrected by the data reduction software and could lead to
inaccuracies in determining continuum properties. Moreover, the
sensitivity of the GIS detectors falls off sharply around 1~\kev. To
deal with these issues we adopted the following conservative
approach. If an observation was made after 1994 we determined if there
was any discrepancy in fit residuals between the SIS and GIS detectors
below 2~\kev. If there was we used the GIS data down to 1~\kev\ and
the SIS data only where the two agreed. If there was no visible
discrepancy, both the GIS and SIS data was used down to 1~\kev. If
data was taken before late 1994 then the low energy SIS data may be
used, but we made sure it was consistent with the GIS data. In all
cases we used data up to 10~\kev. The only exception to this was \ark,
where we used simultaneous \textit{RXTE} data kindly provided by S.\
Vaughan. This data was processed as in Vaughan \etal\ (1999a), except we
used data from all 5 of the satellite's Proportional Counter Units. 
To avoid calibration and background uncertainties we only fit these data
between 3 and 20~\kev. Finally, we have noticed that the S0 data of
\ark\ are inconsistent in the energy range above 4~\kev\ with the other
four detectors which are in good agreement with each other.  Since
applying even more careful data selection than the normal practice
could not eliminate the problem, we attribute it to some anomaly in
the detector.  We therefore use only the S0 data below 4 keV in this
particular data set.

We used \textsc{xspec}~v11.0 for the spectral fitting.  The grid of
ionised disc models that was computed (Section~\ref{sect:model})
was converted into a tablefile that could be read in by
\textsc{xspec}. In each fit, we let the relative normalisation between
the SIS and the GIS vary, but we fixed the cold absorption at the
Galactic value. Since we primarily fit data with energies greater than
1~\kev, any excess absorption within a particular galaxy will not
greatly affect our results. For each galaxy we fit its spectrum two 
different ways using the ionised disc models. First, we fit the data
using just the model spectrum. Second, we included the
effects of relativistic smearing by blurring the model with the Laor
(1991) kernel for the Kerr metric. This blurring has four parameters:
the inner radius of the emitting annulus ($r_{min}$), the outer radius
of the annulus, the exponent of the surface emissivity law (assumed to
go as $r^{\alpha}$), and the inclination of the disc along the line of
sight. The outer radius was fixed to be at 1000~$r_{g}$, where $r_g =
GM/c^2$ is the gravitational radius, (the actual value of this
parameter will not affect the fit as the blurring effects come from
the inner regions of the annulus), and the inclination angle was fixed
at 30~degrees which is typical for Seyfert 1 galaxies (Nandra \etal\
1997a).  We then fit the data with the models using four different
combinations of $r_{min}$ and $\alpha$:
$(r_{min},\alpha)=(10,-2.0;10,-2.5;6,-2.0;6,-2.5)$, where $r_{min}$ is
given in units of gravitational radii. Finally, using the best fitting model 
of the four considered, we fit the inclination angle. 

When the ionised disc models were blurred it was necessary for
energies outside the response matrix of the detector to be used in
order to avoid effects at the edges of the observed energy range. This
was done by using the `extend' command in \textsc{xspec} to extend the
energy range of the response matrix. As long as edge effects were
avoided, the results of the fits do not depend on how far we extended
the response matrix. 

\begin{figure*}
\begin{minipage}{150mm}
\psrotatefirst
\centerline{
        \psfig{file=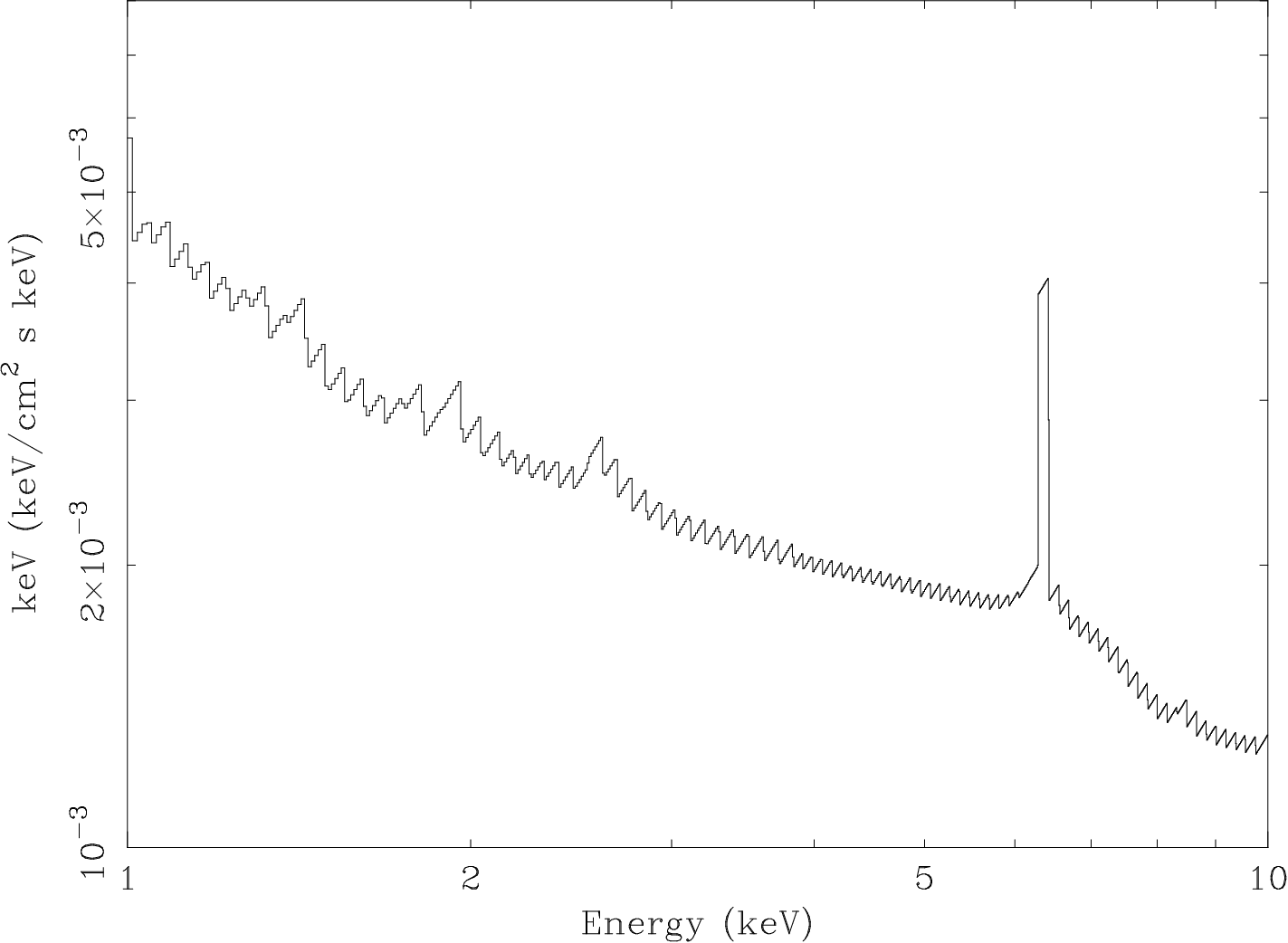,width=3.25in,angle=-90,silent=}
        \psfig{file=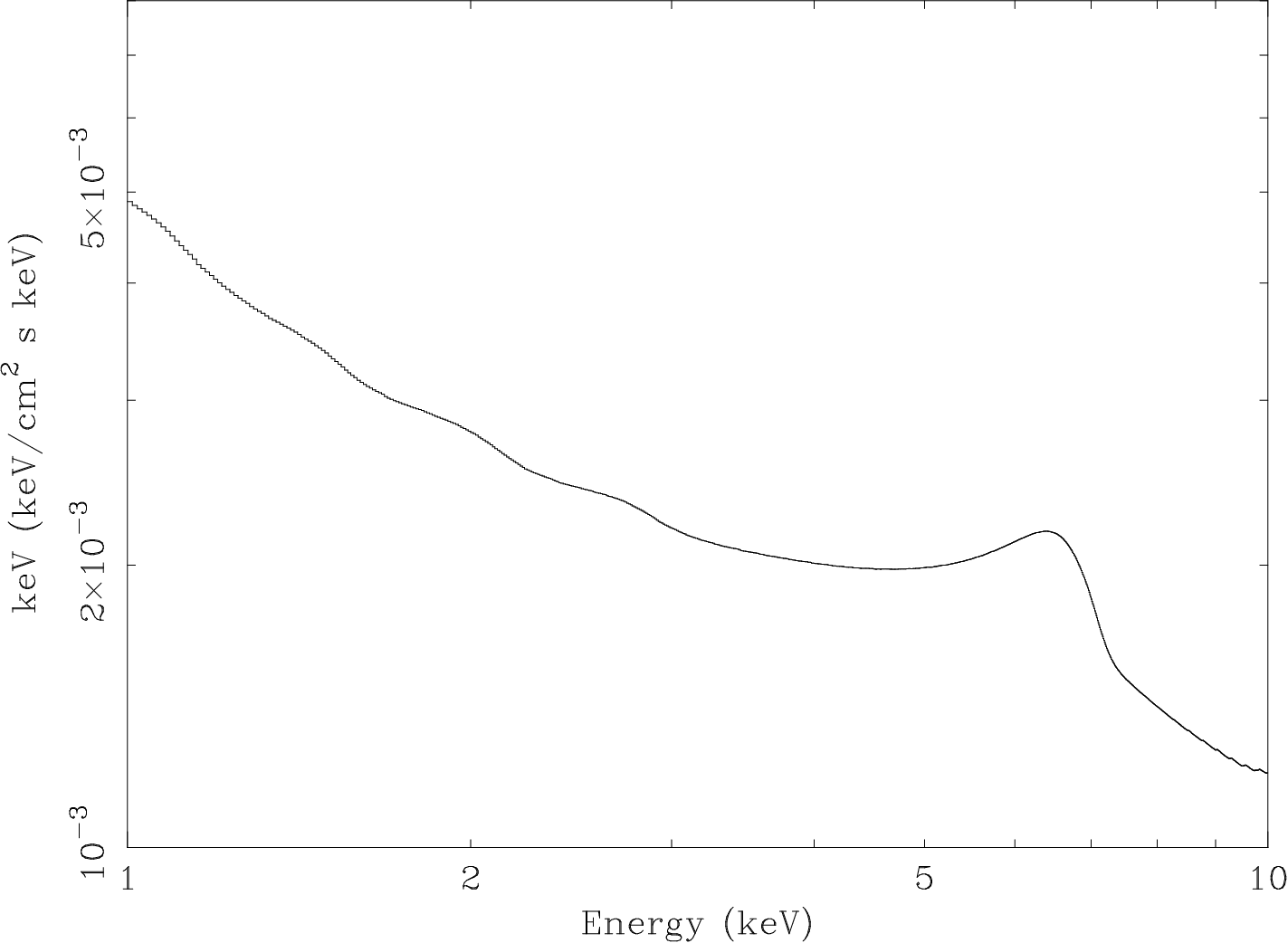,width=3.25in,angle=-90,silent=}
}
\vspace{5mm}
\centerline{
        \psfig{file=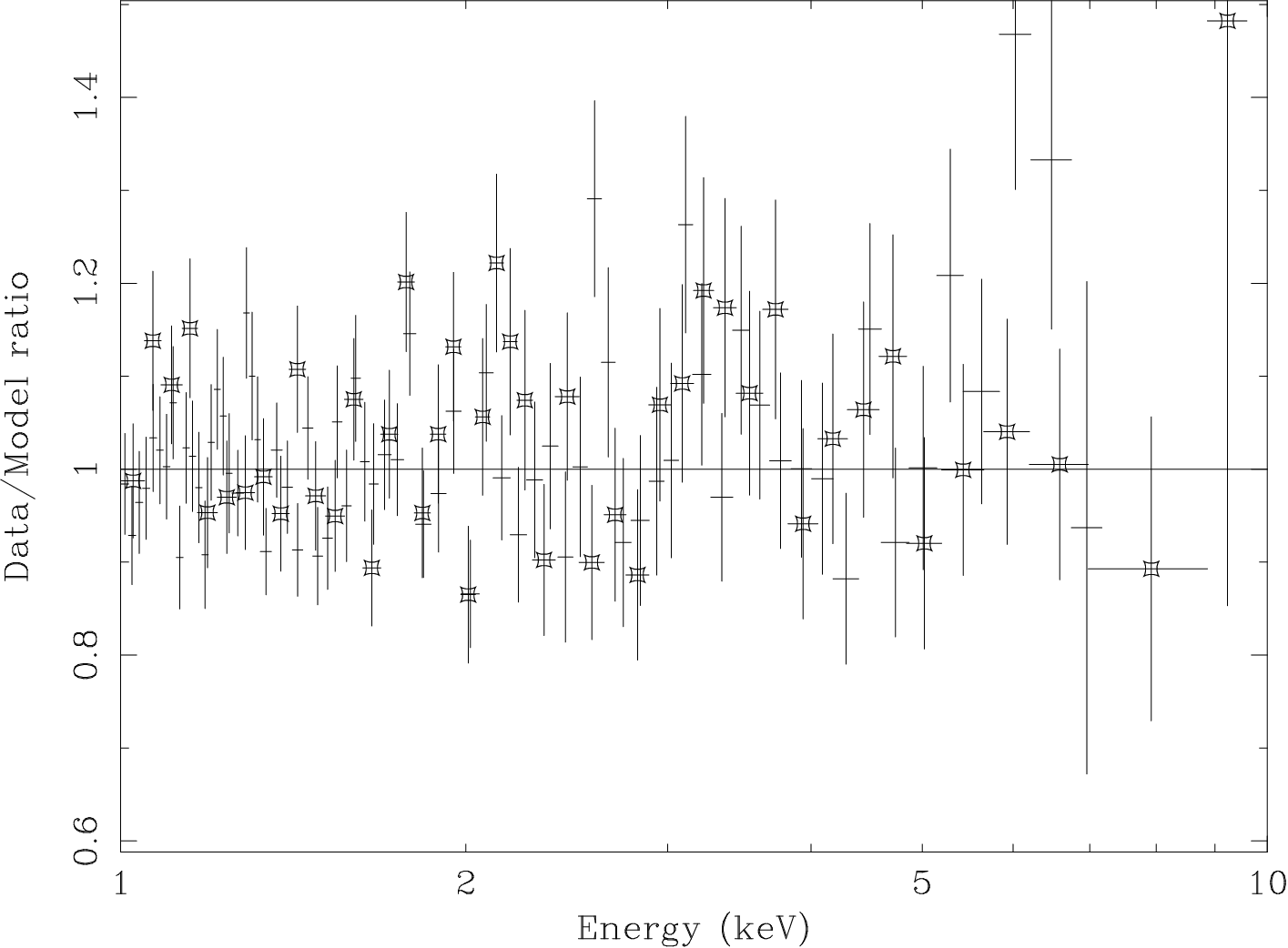,width=3.25in,angle=-90,silent=}
        \psfig{file=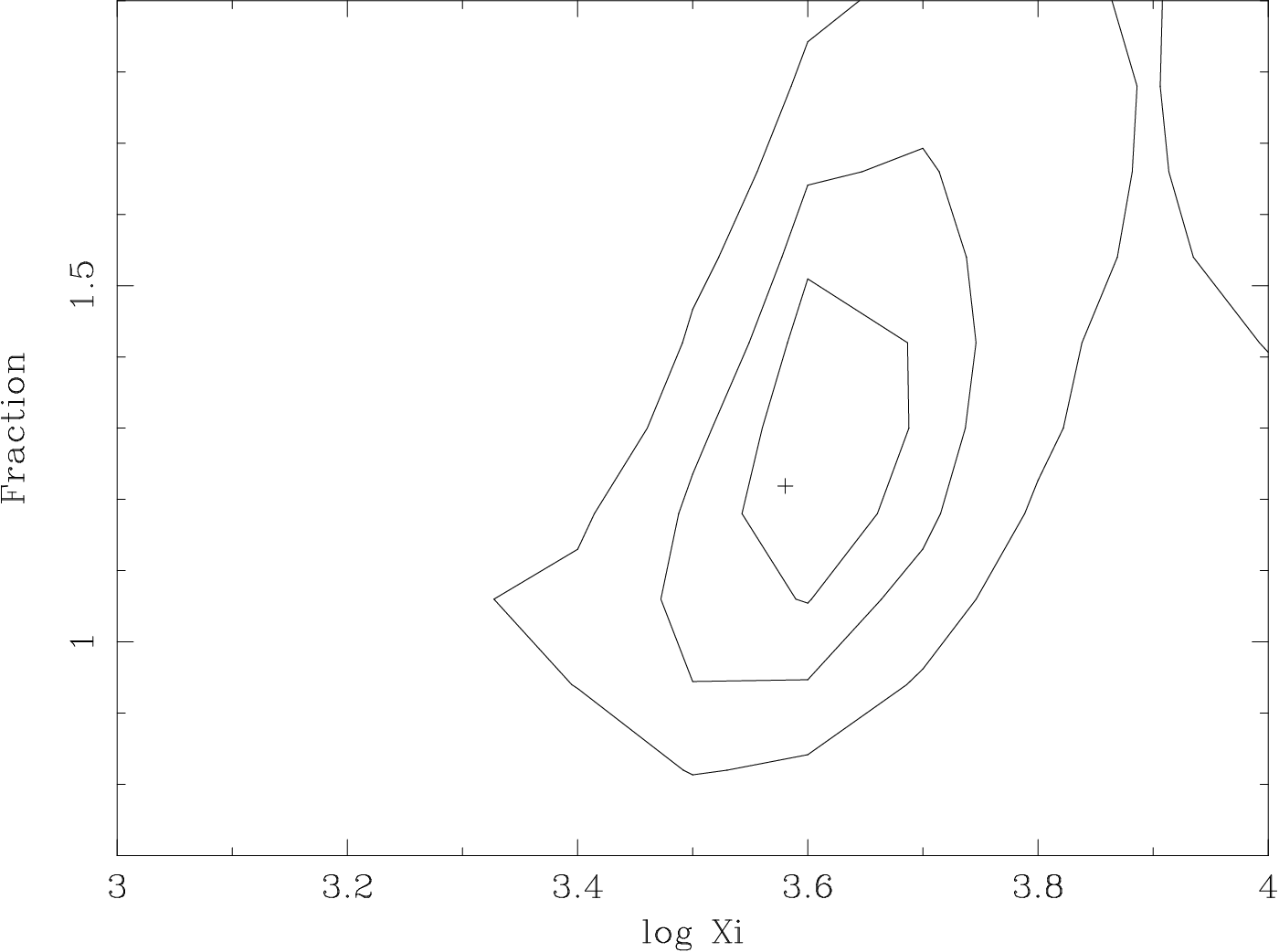,width=3.25in,angle=-90,silent=}
}
\caption{Results from fitting \ton\ with the RF ionised disc model.
The top-left figure shows the best fit unblurred model: $\Gamma= 2.48$,
$\log \xi= 3.30$, and $R= 0.922$. The top-right panel shows the best
fit relativistically blurred model when $i=30$~degrees: $\Gamma=
2.46$, $\log \xi= 3.58$, $R= 1.22$, $r_{min}= 6$~$r_g$, and $\alpha=-2.5$.
The bottom left panel displays the ratio of the data to the best blurred
model from the top right panel. Here, only the S0 data (no symbols) and
the G2 data (open symbols) have been rebinned and plotted.
Finally, the bottom right panel shows the 68, 90 and 99~per~cent confidence 
contours around the best fitting blurred model shown in the top right panel.
The position of the best fit model is denoted by the plus sign.}
\label{fig:ton-results}
\end{minipage}
\end{figure*}

\begin{figure*}
\begin{minipage}{150mm}
\psrotatefirst
\centerline{
        \psfig{file=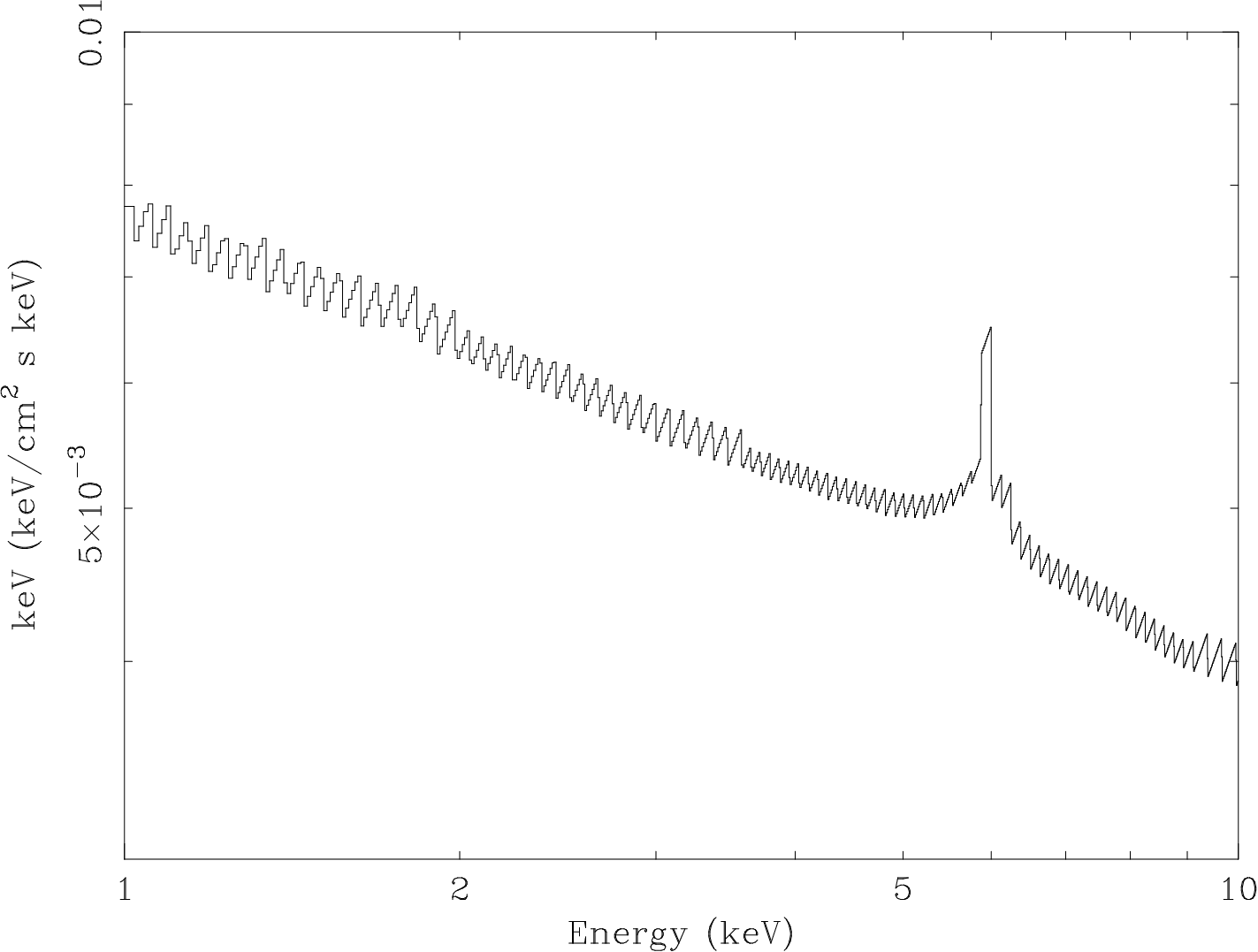,width=3.25in,angle=-90,silent=}
        \psfig{file=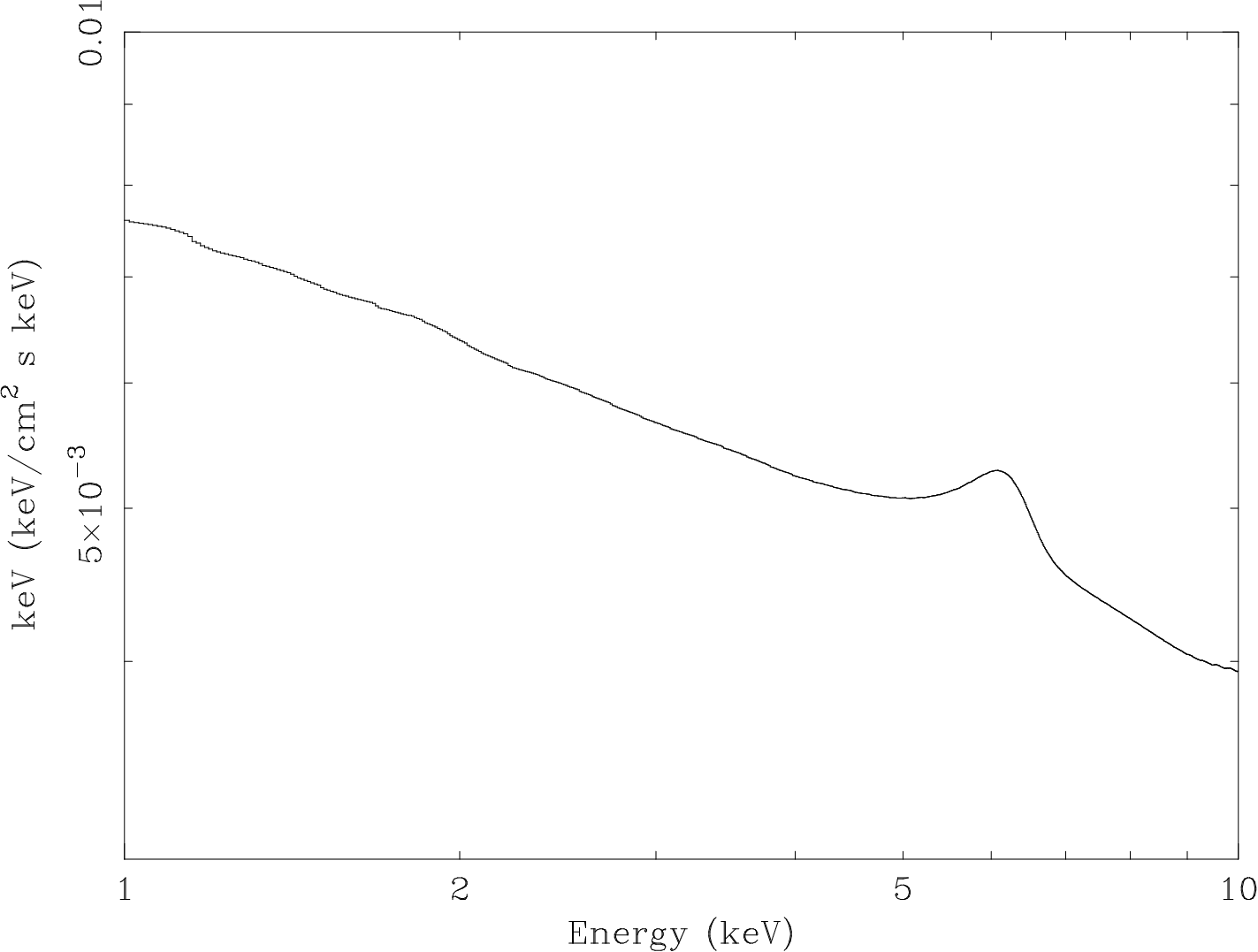,width=3.25in,angle=-90,silent=}
}
\vspace{5mm}
\centerline{
        \psfig{file=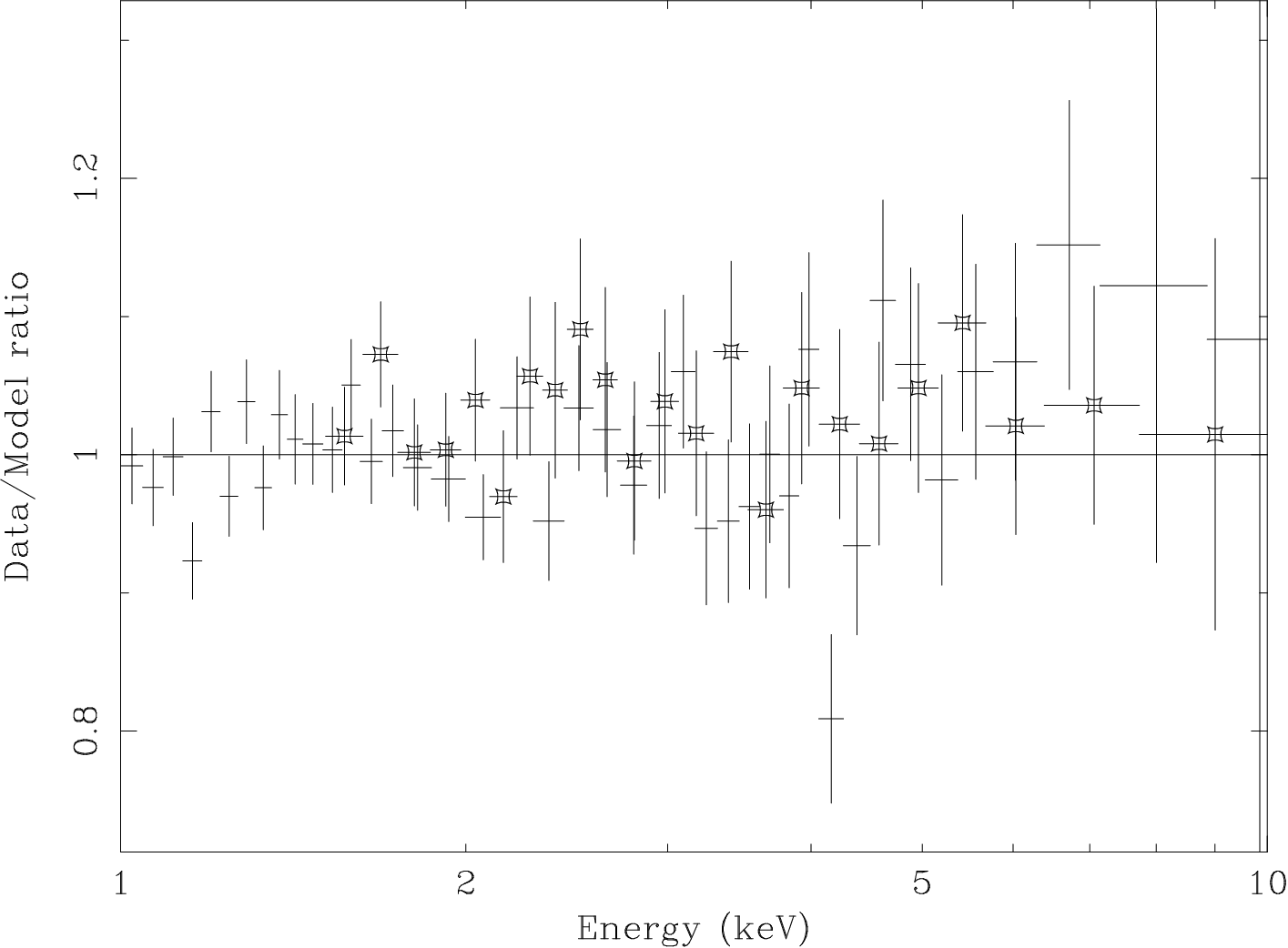,width=3.25in,angle=-90,silent=}
        \psfig{file=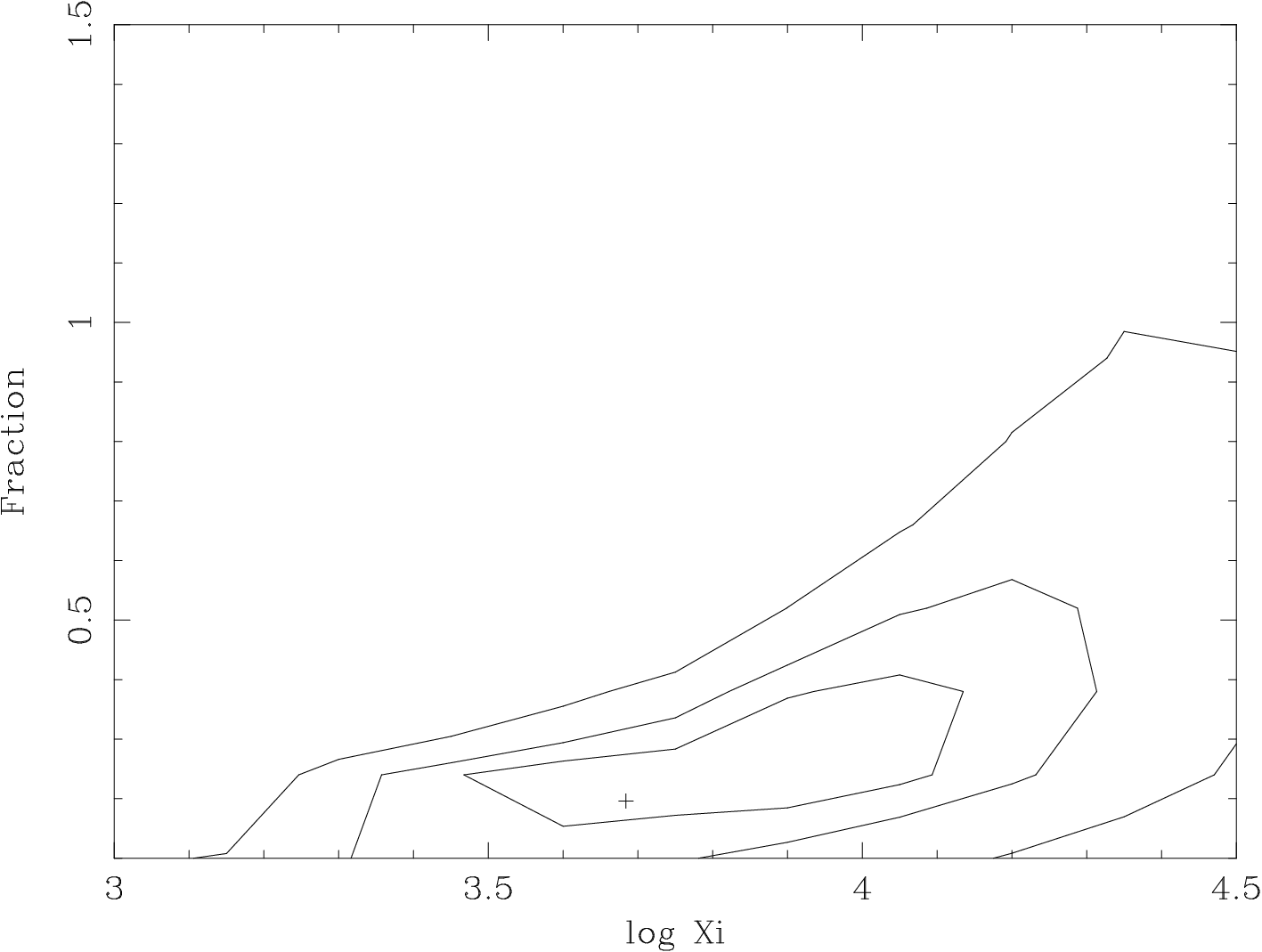,width=3.25in,angle=-90,silent=}
}
\caption{Results from fitting \pks\ with the RF ionised disc model.
The top-left figure shows the best fit unblurred model: $\Gamma= 2.27$,
$\log \xi= 3.69$, and $R= 0.165$. The top-right panel shows the best
fit relativistically blurred model when $i=30$~degrees: $\Gamma=
2.27$, $\log \xi= 3.68$, $R= 0.198$, $r_{min}= 10$~$r_g$, and $\alpha=-2.0$.
The bottom left panel displays the ratio of the data to the best blurred
model from the top right panel. Here, only the S0 data (no symbols) and
the G2 data (open symbols) have been rebinned and plotted.
Finally, the bottom right panel shows the 68, 90 and 99~per~cent confidence 
contours around the best fitting blurred model shown in the top right panel.
The position of the best fit model is denoted by the plus sign.}
\label{fig:pks-results}
\end{minipage}
\end{figure*}

\begin{figure*}
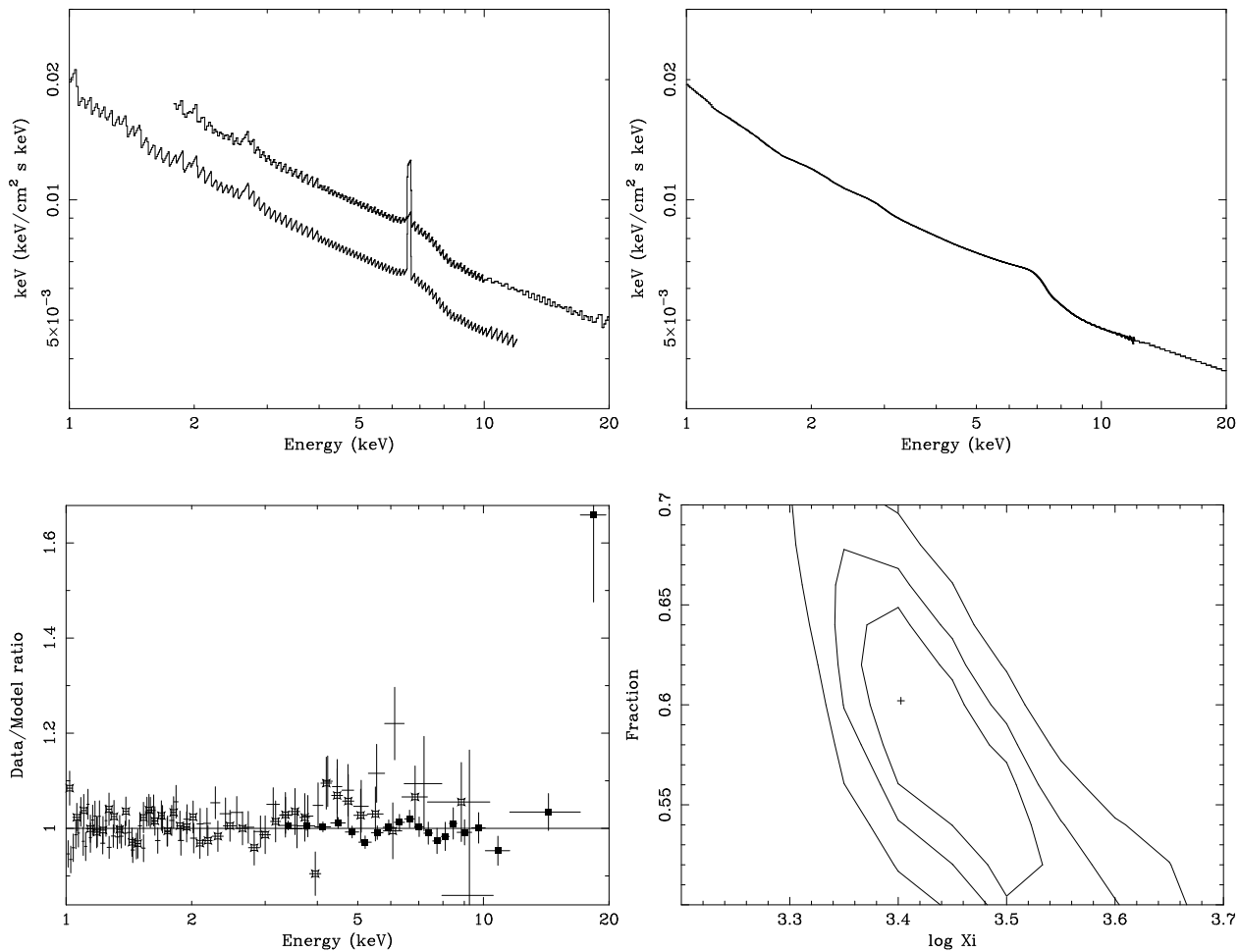

\begin{minipage}{150mm}
\psrotatefirst
\centerline{
        \psfig{file=ark_noblur_no_s0_fp.ps,width=3.25in,angle=-90,silent=}
        \psfig{file=ark_blur_no_s0_fp.ps,width=3.25in,angle=-90,silent=}
}
\vspace{5mm}
\centerline{
        \psfig{file=ark_ratio_no_s0_fp.ps,width=3.25in,angle=-90,silent=}
        \psfig{file=ark_contour_fine_no_s0.ps,width=3.25in,angle=-90,silent=}
}
\caption{Results from fitting \ark\ with the RF ionised disc model.
The top-left figure shows the best fit unblurred model: $\Gamma= 2.59$,
$\log \xi= 3.38$, and $R= 0.520$. The upper curve shows the model through
the \textit{RXTE} response matrix, and the lower curve shows the model 
through the \textit{ASCA} response matrix. The top-right panel shows the best
fit relativistically blurred model when $i=30$~degrees: $\Gamma=
2.59$, $\log \xi= 3.40$, $R= 0.602$, $r_{min}= 6$~$r_g$, and $\alpha=-2.5$.
Here, since the energy range of the response matrices were extended to
do the relativistic blurring, only one curve is displayed.
The bottom left panel displays the ratio of the data to the best blurred
model from the top right panel. Here, the S1 data (no symbols), the
\textit{RXTE} data (solid symbols), and
the G2 data (open symbols) have been rebinned and plotted. Near 20~\kev\
the data is becoming dominated by the background.
Finally, the bottom right panel shows the 68, 90 and 99~per~cent confidence 
contours around the best fitting blurred model shown in the top right panel.
The position of the best fit model is denoted by the plus sign.}
\label{fig:ark-results}
\end{minipage}
\end{figure*}

\begin{figure*}
\begin{minipage}{150mm}
\psrotatefirst
\centerline{
        \psfig{file=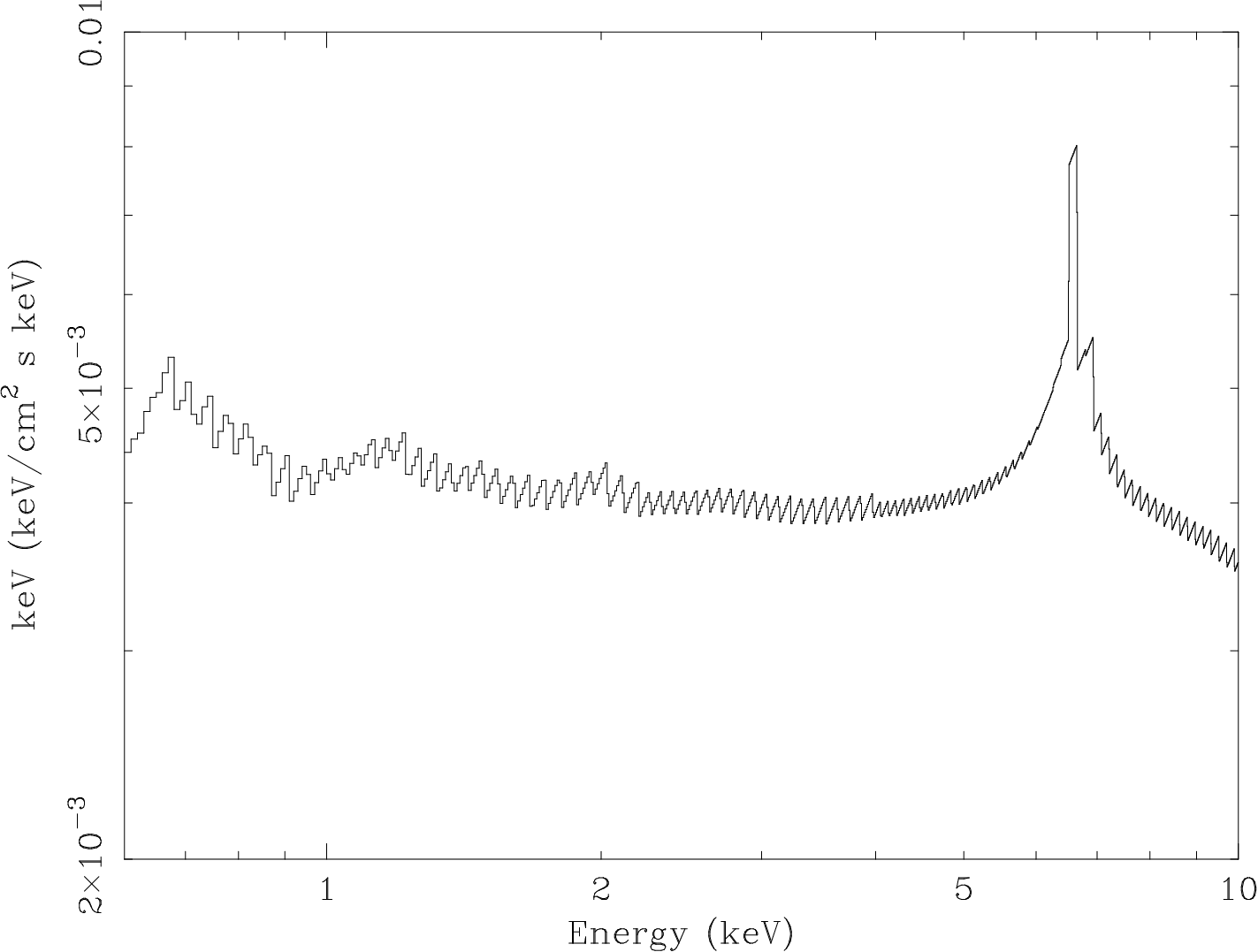,width=3.25in,angle=-90,silent=}
        \psfig{file=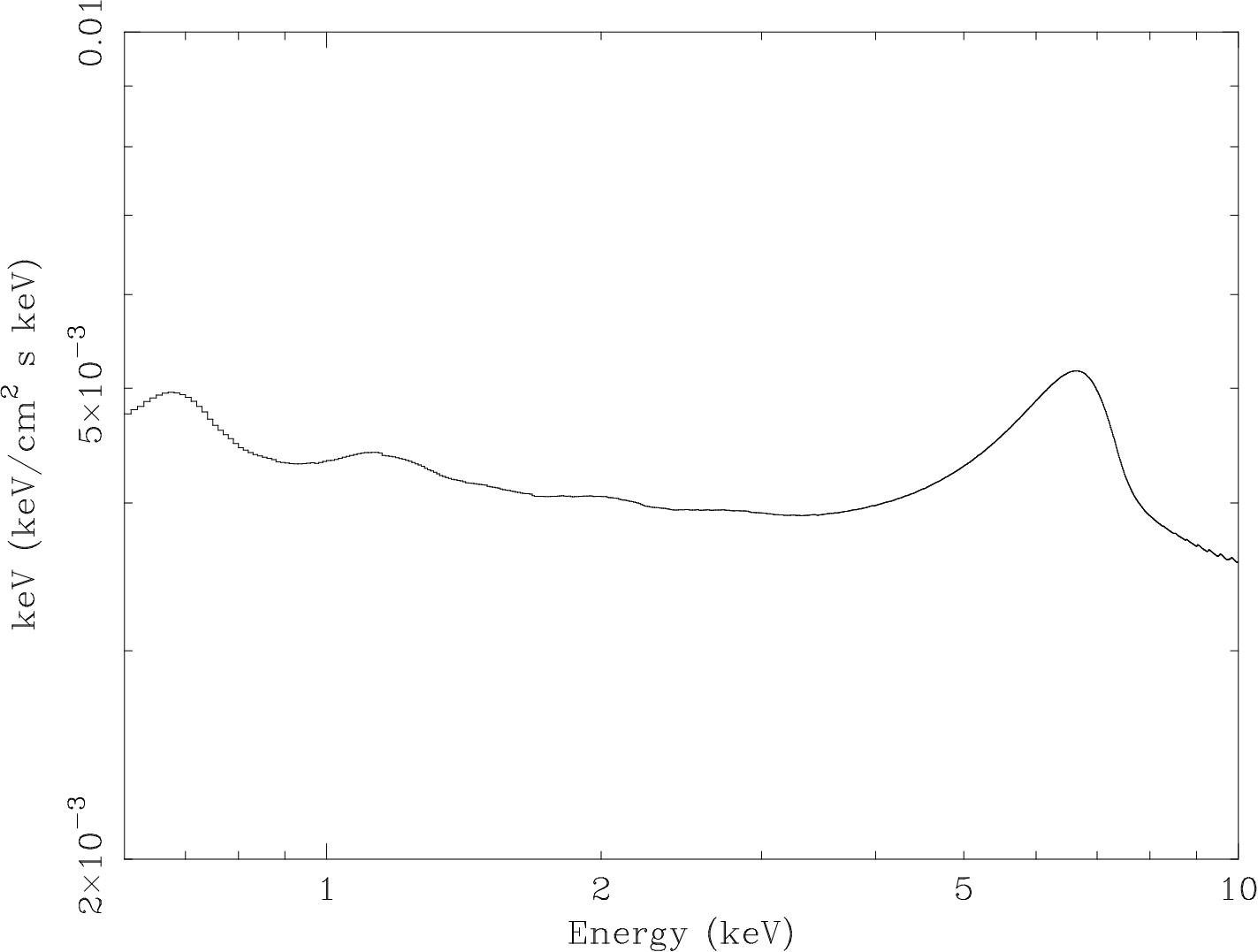,width=3.25in,angle=-90,silent=}
}
\vspace{5mm}
\centerline{
        \psfig{file=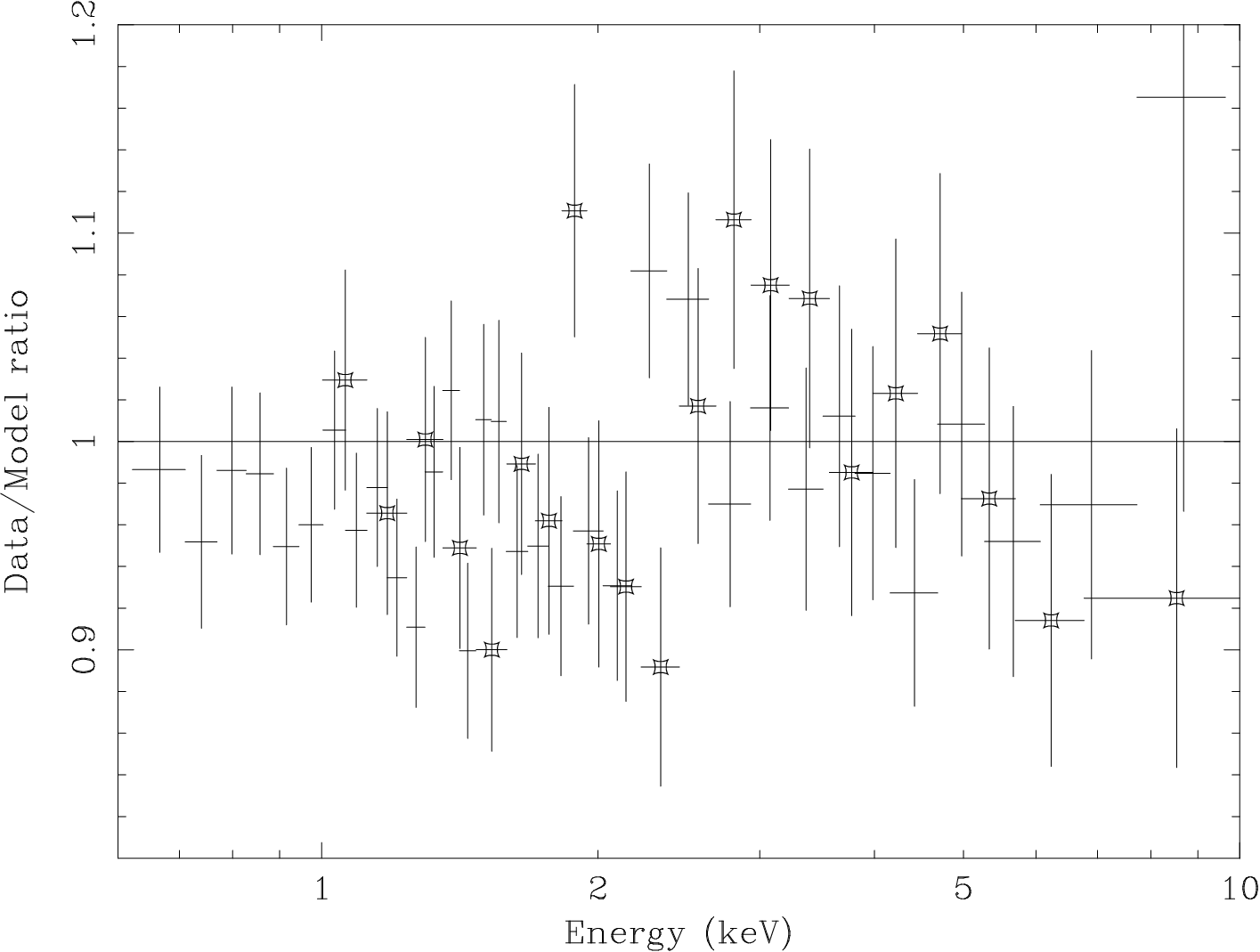,width=3.25in,angle=-90,silent=}
        \psfig{file=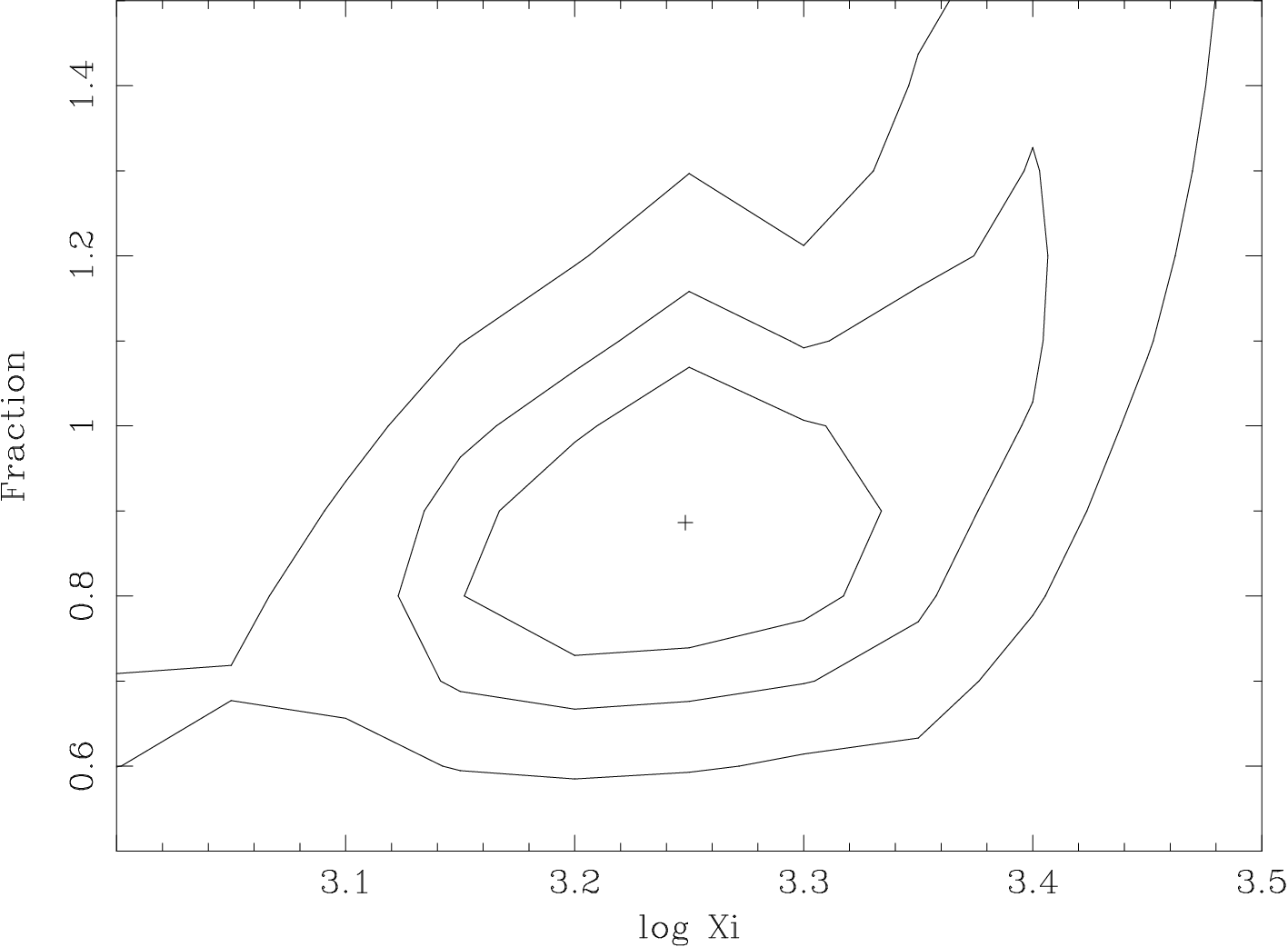,width=3.25in,angle=-90,silent=}
}
\caption{Results from fitting \mrk\ with the RF ionised disc model.
The top-left figure shows the best fit unblurred model: $\Gamma= 1.98$,
$\log \xi= 3.34$, and $R= 0.678$. The top-right panel shows the best
fit relativistically blurred model when $i=30$~degrees: $\Gamma=
1.98$, $\log \xi= 3.25$, $R= 0.888$, $r_{min}= 6$~$r_g$, and $\alpha=-2.5$.
The bottom left panel displays the ratio of the data to the best blurred
model from the top right panel. Here, only the S0 data (no symbols) and
the G2 data (open symbols) have been rebinned and plotted.
Finally, the bottom right panel shows the 68, 90 and 99~per~cent confidence 
contours around the best fitting blurred model shown in the top right panel.
The position of the best fit model is denoted by the plus sign.}
\label{fig:mrk-results}
\end{minipage}
\end{figure*}

\begin{figure*}
\begin{minipage}{150mm}
\psrotatefirst
\centerline{
        \psfig{file=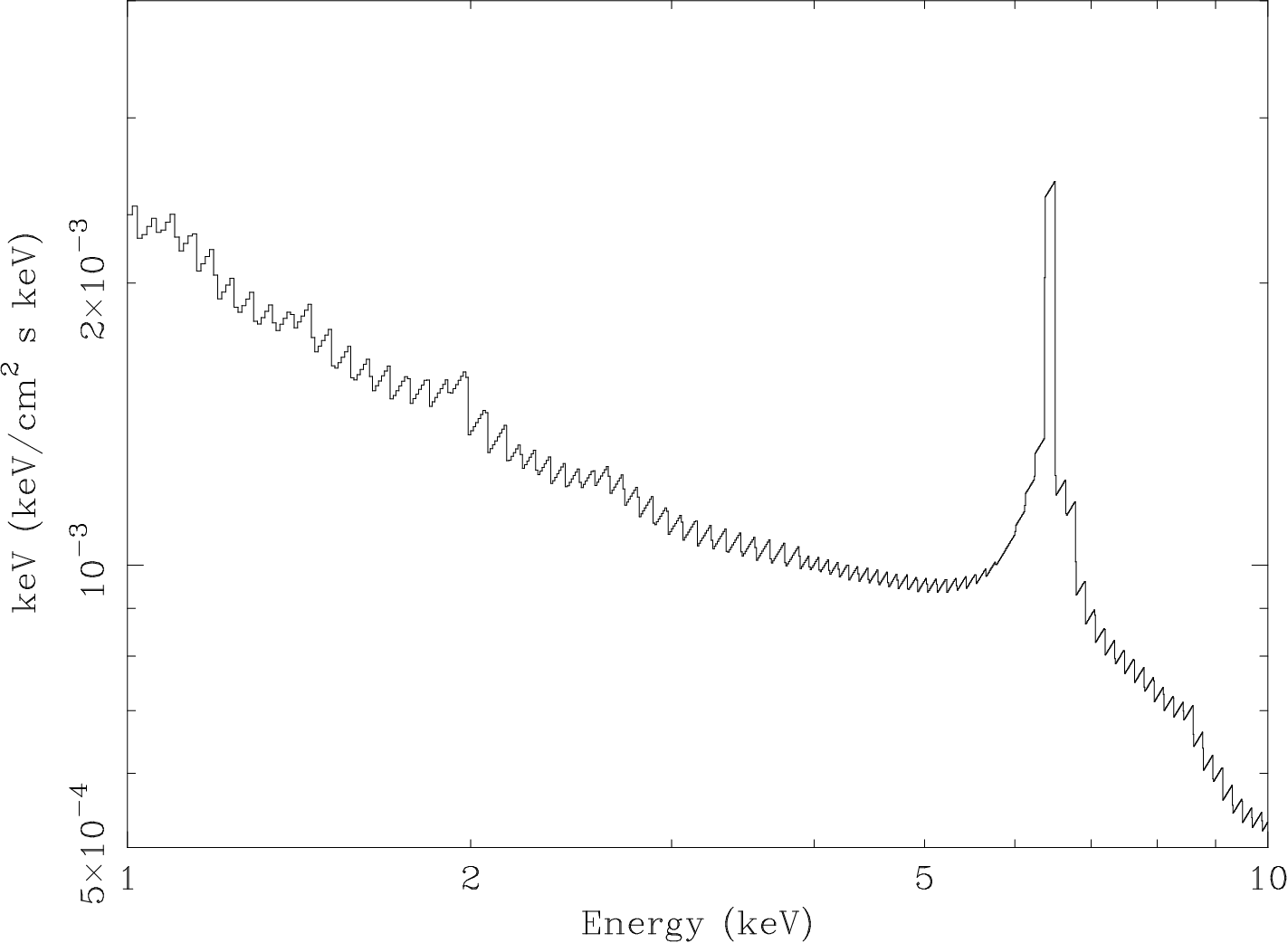,width=3.25in,angle=-90,silent=}
        \psfig{file=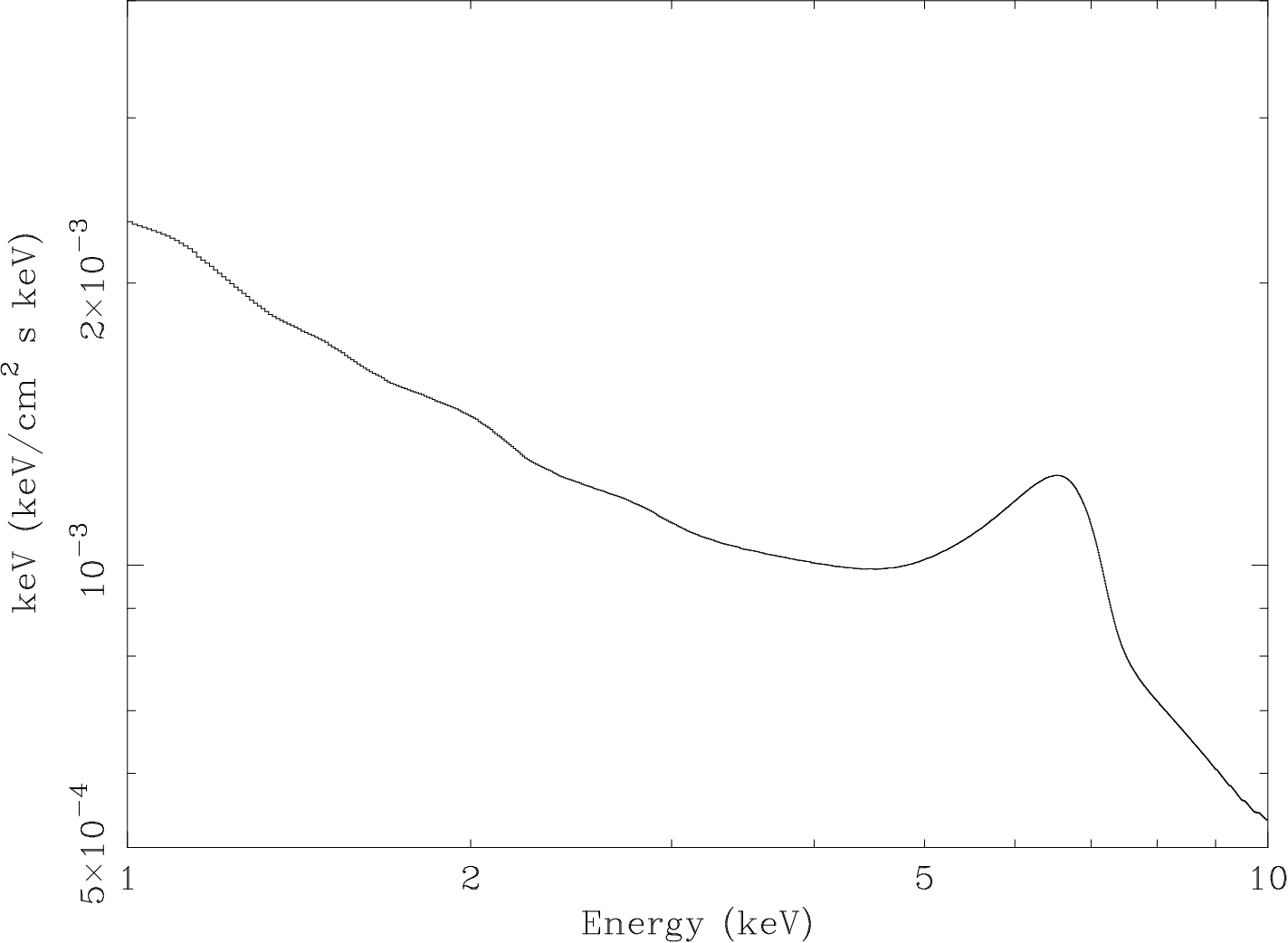,width=3.25in,angle=-90,silent=}
}
\vspace{5mm}
\centerline{
        \psfig{file=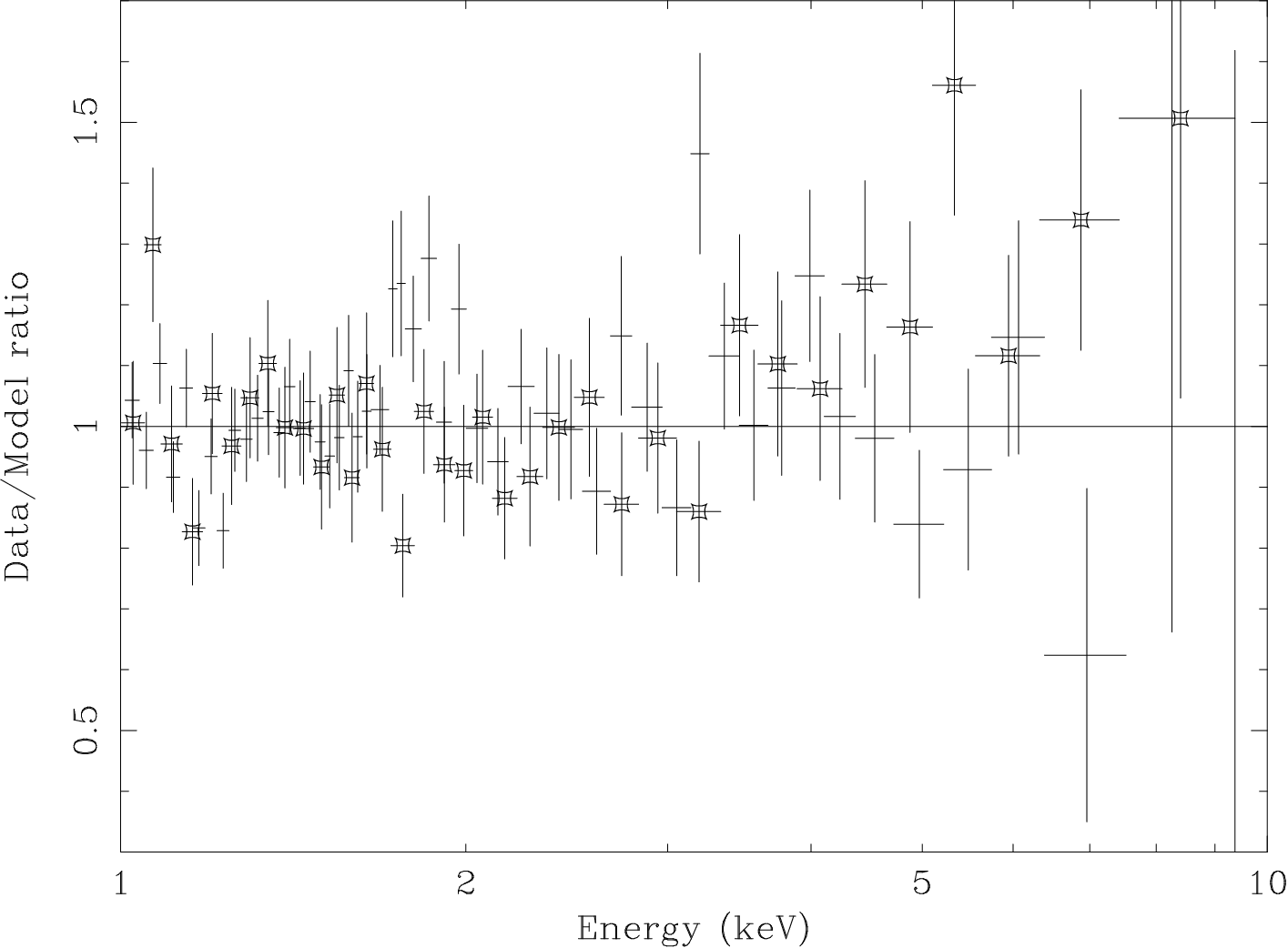,width=3.25in,angle=-90,silent=}
        \psfig{file=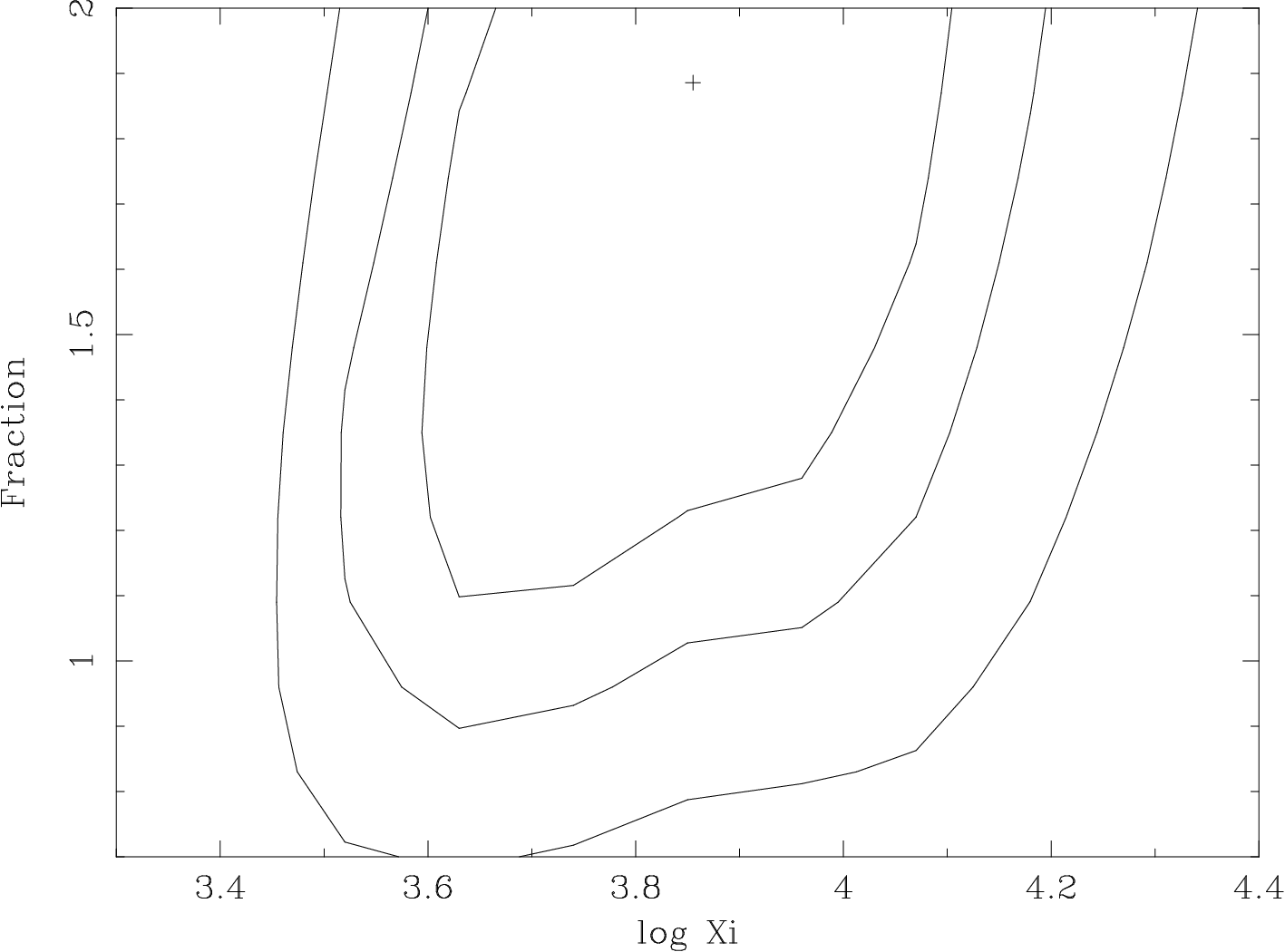,width=3.25in,angle=-90,silent=}
}
\caption{Results from fitting \pg\ with the RF ionised disc model.
The top-left figure shows the best fit unblurred model: $\Gamma= 2.43$,
$\log \xi= 3.91$, and $R= 1.44$. The top-right panel shows the best
fit relativistically blurred model when $i=30$~degrees: $\Gamma=
2.42$, $\log \xi= 3.86$, $R= 1.89$, $r_{min}=10$~$r_g$, and $\alpha=-2.5$.
The bottom left panel displays the ratio of the data to the best blurred
model from the top right panel. Here, only the S0 data (no symbols) and
the G2 data (open symbols) have been rebinned and plotted.
Finally, the bottom right panel shows the 68, 90 and 99~per~cent confidence 
contours around the best fitting blurred model shown in the top right panel.
The position of the best fit model is denoted by the plus sign.}
\label{fig:pg-results}
\end{minipage}
\end{figure*}

The results of these fits are shown in
Table~\ref{table:ross}
\begin{table*}
\begin{minipage}{180mm}
\caption{Best fit parameters of the RF ionised disc models. All values are
quoted in the rest frame. $\Gamma$ is
the photon index, $\xi$ is the ionisation parameter as defined in 
Equation~\ref{eq:ip}, $R$ is the reflection fraction, $r_{min}$ is
the inner radius of the emitting annulus in units of $r_g$, $\alpha$
is the exponent for the disc emissivity law, and $i$ is the inclination
of the disc to the line of sight in degrees.}
\label{table:ross}
\begin{tabular}{@{}lcccccccccc}
Galaxy & Model & Energy & $\Gamma$ & $\log \xi$ & $R$ & $r_{min}$ & $\alpha$ & $i$ & $\chi^2$/dof & $\chi^2_{red}$ \\
 & & Range$^a$ & & & & & & & & \\ \hline
\ton & not blurred & 1--10 & 2.48$\pm 0.02$ & 3.30$_{-0.07}^{+0.33}$ & 0.922$_{-0.221}^{+0.132}$ & & & & 834/860 & 0.970 \\
 & blurred & & 2.44 & 3.54 & 1.041 & 10.0 & $-2.0$ & 30.0$^f$ & 829/860 & 0.964 \\ 
 & & & 2.45 & 3.55 & 1.136 & 10.0 & $-2.5$ & 30.0$^f$ & 824/860 & 0.958 \\
 & & & 2.44 & 3.55 & 1.096 & 6.0 & $-2.0$ & 30.0$^f$ & 827/860 & 0.962 \\
 & & & 2.46$\pm 0.02$ & 3.58$_{-0.07}^{+0.13}$ & 1.22$_{-0.24}^{+0.32}$ & 6.0 & $-2.5$ & 30.0$^f$ & 823/860 & 0.957 \\
 & & & 2.48$_{-0.03}^{+0.02}$ & 3.60$_{-0.09}^{+0.11}$ & 1.205$_{-0.239}^{+0.264}$ & 6.0 & $-2.5$ & 21.8$_{-16}^{+8.1}$ & 820/859 & 0.955 \\
\pks & not blurred & 1--10 & 2.27$_{-0.01}^{+0.02}$ & 3.69$_{-0.44}^{+0.51}$ & 0.165$_{-0.069}^{+0.113}$ & & & & 1022/1054 & 0.970 \\
 & blurred & & 2.27$\pm 0.02$ & 3.68$_{-0.38}^{+0.51}$ & 0.198$_{-0.083}^{+0.172}$ & 10.0 & $-2.0$ & 30.0$^f$ & 1022/1054 & 0.970 \\
 & & & 2.27 & 3.60 & 0.184 & 10.0 & $-2.5$ & 30.0$^f$ & 1023/1054 & 0.971 \\
 & & & 2.27 & 3.60 & 0.187 & 6.0 & $-2.0$ & 30.0$^f$ & 1023/1054 & 0.970 \\
 & & & 2.27 & 3.60 & 0.167 & 6.0 & $-2.5$ & 30.0$^f$ & 1027/1054 & 0.974 \\
 & & & 2.26$\pm 0.02$ & 3.85$_{-0.12}^{+0.23}$ & 0.431$_{-0.162}^{+0.183}$ & 10.0 & $-2.0$ & 89.4$_{-35.5}^{+0.6p}$ & 1016/1053 & 0.965 \\
 & not blurred & & 2.25$\pm 0.02$ & 4.48$_{-0.16}^{+0.27}$ & 1.0$^f$ & & & & 1032/1055 & 0.979 \\
 & blurred & & 2.25$_{-0.01}^{+0.02}$ & 4.48$_{-0.23}^{+0.20}$ & 1.0$^f$ & 10.0 & $-2.0$ & 30.0$^f$ & 1031/1055 & 0.978 \\
 & & & 2.25 & 4.48 & 1.0$^f$ & 10.0 & $-2.5$ & 30.0$^f$ & 1036/1055 & 0.982 \\
 & & & 2.25 & 4.48 & 1.0$^f$ & 6.0 & $-2.0$ & 30.0$^f$ & 1033/1055 & 0.980 \\
 & & & 2.25 & 4.48 & 1.0$^f$ & 6.0 & $-2.5$ & 30.0$^f$ & 1042/1055 & 0.988 \\
 & & & 2.25$_{-0.02}^{+0.01}$ & 4.17$_{-0.15}^{+0.19}$ & 1.0$^f$ & 10.0 & $-2.0$ & 89.9$_{-28.3}^{+0.1p}$ & 1019/1054 & 0.966 \\
\ark & not blurred & 1--20 & 2.59$\pm 0.01$ & 3.38$_{-0.04}^{+0.13}$ & 0.520$_{-0.044}^{+0.054}$ & & & & 1399/1337 & 1.046 \\
 & blurred & & 2.59 & 3.39 & 0.530 & 10.0 & $-2.0$ & 30.0$^f$ & 1377/1337 & 1.030 \\
 & & & 2.59 & 3.40 & 0.564 & 10.0 & $-2.5$ & 30.0$^f$ & 1368/1337 & 1.023 \\
 & & & 2.59 & 3.40 & 0.540 & 6.0 & $-2.0$ & 30.0$^f$ & 1374/1337 & 1.028 \\
 & & & 2.59$\pm 0.01$ & 3.40$_{-0.05}^{+0.15}$ & 0.602$_{-0.103}^{+0.054}$ & 6.0 & $-2.5$ & 30.0$^f$ & 1367/1337 & 1.022 \\
 & & & 2.59$\pm 0.01$ & 3.41$_{-0.05}^{+0.14}$ & 0.607$_{-0.050}^{+0.056}$ & 6.0 & $-2.5$ & 27.4$_{-5.1}^{+2.4}$ & 1364/1336 & 1.021 \\ 
\ngc & not blurred & 0.5--10 & 1.82$\pm 0.01$ & 3.39$\pm 0.01$ & 2.0$_{-0.17}^{+0.0p}$ & & & & 2627/1833 & 1.433 \\
 & blurred & & 1.85 & 3.18 & 1.02 & 10.0 & $-2.0$ & 30.0$^f$ & 2589/1833 & 1.412 \\
 & & & 1.85 & 3.19 & 1.10 & 10.0 & $-2.5$ & 30.0$^f$ & 2445/1833 & 1.334 \\
 & & & 1.85 & 3.18 & 1.04 & 6.0 & $-2.0$ & 30.0$^f$ & 2528/1833 & 1.379 \\
 & & & 1.85$\pm 0.01$ & 3.198$\pm 0.004$ & 1.21$_{-0.06}^{+0.08}$ & 6.0 & $-2.5$ & 30.0$^f$ & 2349/1833 & 1.281 \\
 & & & 1.86$\pm 0.01$ & 3.200$_{-0.003}^{+0.004}$ & 1.19$_{-0.05}^{+0.07}$ & 6.0 & $-2.5$ & 21.6$_{-2.9}^{+3.0}$ & 2351/1832 & 1.283 \\ 
\mrk & not blurred & 0.6--10 & 1.98$\pm 0.02$ & 3.34$_{-0.07}^{+0.06}$ & 0.678$_{-0.137}^{+0.217}$ & & & & 800/733 & 1.091 \\
 & blurred & & 1.98 & 3.27 & 0.763 & 10.0 & $-2.0$ & 30.0$^f$ & 797/733 & 1.087 \\
 & & & 1.98 & 3.26 & 0.836 & 10.0 & $-2.5$ & 30.0$^f$ & 788/733 & 1.076 \\
 & & & 1.98 & 3.26 & 0.792 & 6.0 & $-2.0$ & 30.0$^f$ & 793/733 & 1.082 \\
 & & & 1.98$\pm 0.02$ & 3.25$\pm 0.10$ & 0.888$_{-0.180}^{+0.197}$ & 6.0 & $-2.5$ & 30.0$^f$ & 785/733 & 1.071 \\
 & & & 1.99$\pm 0.02$ & 3.28$_{-0.12}^{+0.11}$ & 0.876$_{-0.165}^{+0.112}$ & 6.0 & $-2.5$ & 21.0$_{-11.9}^{+8.5}$ & 782/732 & 1.068 \\
\pg & not blurred & 1--10 & 2.43$\pm 0.04$ & 3.91$_{-0.39}^{+0.28}$ & 1.44$_{-0.48}^{+0.56p}$ & & & & 544/517 & 1.052 \\
 & blurred & & 2.41 & 3.86 & 1.83 & 10.0 & $-2.0$ & 30.0$^f$ & 544/517 & 1.052 \\
 & & & 2.42$_{-0.03}^{+0.04}$ & 3.86$_{-0.29}^{+0.26}$ & 1.89$_{-0.69}^{+0.11p}$ & 10.0 & $-2.5$ & 30.0$^f$ & 543/517 & 1.050 \\
 & & & 2.41 & 3.85 & 1.89 & 6.0 & $-2.0$ & 30.0$^f$ & 543/517 & 1.051 \\
 & & & 2.43 & 3.85 & 1.88 & 6.0 & $-2.5$ & 30.0$^f$ & 544/517 & 1.052 \\
 & & & 2.44$\pm 0.04$ & 3.76$_{-0.19}^{+0.37}$ & 1.46$_{-0.48}^{+0.54p}$ & 10.0 & $-2.5$ & 15.2$_{-15.2p}^{+13.6}$ & 540/516 & 1.046 \\
\end{tabular}
\medskip

$^a$ units of keV 

$^f$ parameter fixed at value

$^p$/$_p$ parameter pegged at upper/lower limit
\end{minipage}
\end{table*} 
and in Figures~\ref{fig:ton-results}--\ref{fig:pg-results}. Plots of
the ratio of the data to a simple absorbed power-law model can be seen
elsewhere (Leighly 1999b). Note that all the spectra in 
Figs.~\ref{fig:ton-results}--\ref{fig:pg-results} are plotted in
the observed frame, while all the fit parameters are quoted in the
rest frame. The errors reported in Table~\ref{table:ross} and in all
subsequent tables correspond to a 90~per~cent confidence interval for
one interesting parameter.

\begin{figure*}
\begin{minipage}{150mm}
\psrotatefirst
\centerline{
        \psfig{file=ngc_noblur_fp.ps,width=3.25in,angle=-90,silent=}
        \psfig{file=ngc_blur_fp.ps,width=3.25in,angle=-90,silent=}
}
\vspace{5mm}
\centerline{
        \psfig{file=ngc_ratio_fp.ps,width=3.25in,angle=-90,silent=}
        \psfig{file=ngc_contour_fine.ps,width=3.25in,angle=-90,silent=}
}
\caption{Results from fitting \ngc\ with the RF ionised disc model.
The top-left figure shows the best fit unblurred model: $\Gamma= 1.82$,
$\log \xi= 3.39$, and $R= 2.0$. The top-right panel shows the best
fit relativistically blurred model when $i=30$~degrees: $\Gamma=
1.85$, $\log \xi= 3.198$, $R= 1.21$, $r_{min}= 6$~$r_g$, and $\alpha=-2.5$.
The bottom left panel displays the ratio of the data to the best blurred
model from the top right panel. Here, only the S0 data (no symbols) and
the G2 data (open symbols) have been rebinned and plotted.
Finally, the bottom right panel shows the 68, 90 and 99~per~cent confidence 
contours around the best fitting blurred model shown in the top right panel.
The position of the best fit model is denoted by the plus sign.}
\label{fig:ngc-results}
\end{minipage}
\end{figure*}

To determine the \fe\ line parameters of the galaxies, we fit the 
spectra above 3~\kev\ with a simple absorbed power-law model. To
obtain a good estimate of the photon index, $\Gamma$, we
ignored the data between 5 and 7~\kev\ (observed frame) to avoid
the influence of the \fe\ line. Once we obtained the best-fitting
value of $\Gamma$, we fixed it at that value and fit a Gaussian
profile to the \fe\ line. The energy centroid of the Gaussian was
constrained to lie between 5.5 and 7.5~\kev, and its width ($\sigma$)
between 0 and 2~\kev. The rest frame fit parameters are shown in 
Table~\ref{table:feline}.
\begin{table*}
\begin{minipage}{140mm}
\caption{Best fit \fe\ line parameters of the sample galaxies. All values
are quoted in the rest frame.
$\Gamma$ is the photon index, $E_{Fe}$ is the energy centroid
of the Gaussian fit in \kev, $\sigma$ is the width of the Gaussian
also in \kev\, and EW is the equivalent width of the line measured
in eV.}
\label{table:feline}
\begin{tabular}{@{}lcccccccc}
Galaxy & Model & Energy & $\Gamma$ & $E_{Fe}$ & $\sigma$ & EW$$ & $\chi^2$/dof & $\chi^2_{red}$ \\
 & & Range$^a$ & & & & & & \\ \hline
\ton & plaw & 3--5, 7--10 & 2.44$\pm 0.14$ & & & & 207/234 & 0.886 \\
 & plaw+gauss & 3--10 & 2.44$^f$ & 6.58$_{-0.19}^{+0.16}$ & 0.382$_{-0.152}^{+0.264}$ & 435 & 275/318 & 0.864 \\
\pks & plaw & 3--5, 7--10 & 2.24$\pm 0.09$ & & & & 362/403 & 0.899 \\
 & plaw+gauss & 3--10 & 2.24$^f$ & 7.34$_{-0.59}^{+0.16p}$ & 0.985$_{-0.327}^{+0.965}$ & 347 & 495/542 & 0.913 \\
\ark & plaw & 3--5, 7--20 & 2.57$\pm 0.02$ & & & & 559/552 & 1.013 \\
 & plaw+gauss & 3--20 & 2.57$^f$ & 6.50$_{-0.25}^{+0.17}$ & 0.471$_{-0.238}^{+0.252}$ & 174 & 707/723 & 0.978 \\
\ngc & plaw & 3--5, 7--10 & 1.84$\pm 0.04$ & & & & 827/735 & 1.125 \\
 & plaw+gauss & 3--10 & 1.84$^f$ & 6.28$_{-0.11}^{+0.10}$ & 0.407$_{-0.227}^{+0.173}$ & 255 & 1205/1129 & 1.067 \\
\mrk & plaw & 3--5, 7--10 & 1.92$_{-0.12}^{+0.10}$ & & & & 203/217 & 0.934 \\
 & plaw+gauss & 3--10 & 1.92$^f$ & 6.43$_{-0.10}^{+0.12}$ & 0.142$_{-0.141}^{+0.174}$ & 189 & 282/303 & 0.929 \\
\pg & plaw & 3--5, 7--10 & 2.26$_{-0.22}^{+0.29}$ & & & & 116/123 & 0.936 \\
 & plaw+gauss & 3--10 & 2.26$^f$ & 6.66$_{-0.44}^{+0.40}$ & 0.556$_{-0.448}^{+0.606}$ & 594 & 165/171 & 0.964 \\
\end{tabular}
\medskip

$^a$: units of keV 

$^f$: parameter fixed at value

$^p$/$_p$: parameter pegged at upper/lower limit
\end{minipage}
\end{table*} 
Despite the differences in data reduction and analysis methodology, our
values compare well with previously published results (e.g., Leighly 1999b; 
Vaughan \etal\ 1999b). 

As a check on the relative importance of our ionised disc fits we also
fit the spectra with three alternative models: a simple power-law plus 
gaussian line model, the warm absorber \textsc{absori} model 
(Done \etal\ 1992; Zdziarski \etal\ 1995), and the ionised disc \textsc{pexriv}
model. The latter two models are included in the \textsc{xspec}
package. The \fe\ line parameters that were
found earlier were included in these fits but were fixed at their
best-fitting values. The results of these fits are shown in
Table~\ref{table:altmodels}. The displayed values of $\xi$ have been
corected for the different energy range in their definition, and so
can be compared with the values in Table~\ref{table:ross}.
\begin{table*}
\begin{minipage}{180mm}
\caption{Fit parameters from the alternative models considered. All values
are quoted in the rest frame. $\Gamma$ is
the photon index and EW is the equivalent width of the \fe\ line in eV (the 
line energy and width are fixed with the values shown in 
Table~\ref{table:feline}). $\xi$ is the ionisation parameter (in units of
\ergcms\ ) of the warm absorber (for the \textsc{absori} models) or the 
ionised disc (for the \textsc{pexriv} models). The value of $\xi$ has been 
corrected to the same energy range as the RF models. $N_H$ is the column 
density of the warm absorber in units of 10$^{22}$~cm$^{-2}$. $R$ is the 
reflection fraction given by the \textsc{pexriv} models. The inclination
angle of the disc has been fixed at 30~degrees for these models.}
\label{table:altmodels}
\begin{tabular}{@{}lccccccccc}
Galaxy & Model & Energy & $\Gamma$ & EW & $\xi$ & $N_H$ & $R$ & $\chi^2$/dof & $\chi^2_{red}$ \\
 & & Range$^a$ & & & & & & & \\ \hline
\ton & plaw+gauss & 1--10 & 2.44$^f$ & 365 & & & & 1204/862 & 1.396 \\ 
 & plaw+\textsc{absori}+gauss & & 2.44$^f$ & 120 & 3180$_{-641}^{+20p}$ & 21.1$_{-5.1}^{+2.6}$ & & 982/860 & 1.142 \\
 & & & 2.64$_{-0.02}^{+0.03}$ & 600 & 1379$_{-336}^{+469}$ & 3.32$_{-0.76}^{+0.89}$ & & 833/859 & 0.970 \\
 & \textsc{pexriv}+gauss & & 2.44$^f$ & 51 & 2794$_{-465}^{+406p}$ & & 1.05$\pm 0.15$ & 907/860 & 1.055 \\
 & & & 2.68$_{-0.06}^{+0.05}$ & 235 & 1041$_{-162}^{+246}$ & & 2.12$_{-0.63}^{+0.86}$ & 821/859 & 0.955 \\
\pks & plaw+gauss & 1--10 & 2.24$^f$ & 148 & & & & 1079/1056 & 1.022 \\ 
 & plaw+\textsc{absori}+gauss & & 2.24$^f$ & 52 & 3200$_{-659}^{+0p}$ & 10.9$_{-4.9}^{+5.2}$ & & 1067/1054 & 1.012 \\
 & & & 2.34$_{-0.03}^{+0.02}$ & 548 & 0.0$_{-0p}^{+3200p}$ & 0.02$_{-0.02p}^{+0.30}$ & & 1012/1053 & 0.961 \\
 & \textsc{pexriv}+gauss & & 2.24$^f$ & 165 & 3199$_{-860}^{+1p}$ & & 0.15$\pm 0.07$ & 1035/1054 & 0.982 \\
 & & & 2.32$\pm 0.03$ & 450 & 0.0$_{-0p}^{+3200p}$ & & 0.50$_{-0.50p}^{+0.62}$ & 1011/1053 & 0.961 \\
\ark & plaw+gauss & 1--20 & 2.57$^f$ & 164 & & & & 2232/1339 & 1.59 \\
 & plaw+\textsc{absori}+gauss & & 2.57$^f$ & 35 & 3200$_{-145}^{+0p}$ & 6.9$_{-0.5}^{+0.6}$ & & 1776/1337 & 1.328 \\
 & & & 2.71$\pm 0.01$ & 233 & 1032$_{-124}^{+212}$ & 1.8$\pm 0.2$ & & 1467/1336 & 1.098 \\
 & \textsc{pexriv}+gauss & & 2.57$^f$ & 0.0 & 3200$_{-184}^{+0.0p}$ & & 0.53$_{-0.06}^{+0.03}$ & 1635/1337 & 1.223 \\
 & & & 2.73$\pm 0.02$ & 0.0 & 968$_{-126}^{+149}$ & & 1.4$\pm 0.2$ & 1416/1336 & 1.060 \\
\ngc & plaw+gauss & 0.5--10 & 1.84$^f$ & 40 & & & & 7713/1835 & 4.203 \\
 & plaw+\textsc{absori}+gauss & & 1.84$^f$ & 0.0 & 1107 & 73 & & 5635/1833 & 3.074 \\
 & & & 2.031$\pm 0.006$ & 316 & 3200$_{-136}^{+0p}$ & 99$\pm 1$ & & 3351/1832 & 1.829 \\
 & \textsc{pexriv}+gauss & & 1.84$^f$ & 0.0 & 668 & & 1.1 & 3945/1833 & 2.152 \\
 & & & 2.11$_{-0.01}^{+0.02}$ & 27 & 83.9$_{-4.6}^{+3.2}$ & & 2.4$_{-0.2}^{+0.3}$ & 2429/1832 & 1.326 \\
\mrk & plaw+gauss & 0.6--10 & 1.92$^f$ & 87 & & & & 1328/735 & 1.807 \\ 
 & plaw+\textsc{absori}+gauss & & 1.92$^f$ & 0.0 & 705$_{-144}^{+181}$ & 22.9$_{-6.6}^{+8.6}$ & & 1183/733 & 1.614 \\
 & & & 2.15$\pm 0.02$ & 246 & 157$_{-33}^{+59}$ & 2.8$_{-0.7}^{+0.8}$ & & 804/732 & 1.099 \\
 & \textsc{pexriv}+gauss & & 1.92$^f$ & 0.0 & 666$_{-117}^{+148}$ & & 0.84$_{-0.12}^{+0.11}$ & 1034/733 & 1.411 \\
 & & & 2.14$\pm 0.04$ & 140 & 138$_{-41}^{+81}$ & & 1.26$_{-0.39}^{+0.58}$ & 792/732 & 1.082 \\
\pg & plaw+gauss & 1--10 & 2.26$^f$ & 59 & & & & 947/519 & 1.824 \\
 & plaw+\textsc{absori}+gauss & & 2.26$^f$ & 0.0 & 3200$_{-108}^{+0p}$ & 77$_{-12}^{+8}$ & & 766/517 & 1.482 \\
 & & & 2.71$_{-0.06}^{+0.08}$ & 1400 & 567$_{-200}^{+594}$ & 1.1$_{-0.55}^{+0.54}$ & & 556/516 & 1.077 \\
 & \textsc{pexriv}+gauss & & 2.26$^f$ & 0.0 & 3199$_{-295}^{+1p}$ & & 1.8$_{-0.3}^{+0.5}$ & 723/517 & 1.399 \\
 & & & 2.80$_{-0.07}^{+0.09}$ & 336 & 530$_{-355}^{+219}$ & & 4.1$_{-1.4}^{+3.0}$ & 540/516  & 1.046 \\
\end{tabular}
\medskip
 
$^a$: units of keV 

$^f$: parameter fixed at value

$^p$/$_p$: parameter pegged at upper/lower limit
\end{minipage}
\end{table*}
The \textsc{absori} models were run with the temperature of the
absorber fixed at 10$^6$~K, and with solar Fe abundance (adjusted to
match the abundances used in the RF models). Similarly, the
\textsc{pexriv} models had the disc temperature fixed at 10$^6$~K, the
inclination angle fixed at 30 degrees, and solar abundances (after
correction). The power-law cutoff energy for the \textsc{pexriv} models was
fixed at 100~\kev, in agreement with our models.

\subsection{Interpretation of the spectral fitting results}
\label{sub:res-interp}
Studying the results listed in 
Tables~\ref{table:ross}--\ref{table:altmodels} gives rise to the
following comments:
\begin{enumerate}
\item The RF ionised disc model provides an adequate fit to the 
\textit{ASCA} data in five out of the six sources examined. The
only source that the model does not fit is \ngc, which contains a
warm absorber and is well known for its spectral complexity (Guainazzi 
\etal\ 1996). 
An attempt to fully account for all the features in the spectrum of
\ngc\ is beyond the scope of this paper. 

\item In general, the alternative models presented in 
Table~\ref{table:altmodels} have difficulty fitting the data with
realisitic values of the parameters. These models seem to be able 
to fit the data only with a large value of the photon index, $\Gamma$.
If $\Gamma$ is fixed at the value determined from the 3--10~\kev\ fits,
then the fits are quite poor and often results in an extremely small \fe\
 EW. This is because of the unrealistically strong Fe K edge one finds
in the \textsc{absori} and \textsc{pexriv} models (e.g, Ross \etal\ 1999).
The RF models determine a value of $\Gamma$ that is comparable with the 
ones determined from the high energy fit, and so does a good job matching
the intrinsic continuum shape. 

\item The RF results show that the accretion discs in these sources
might be relatively highly ionised with $\xi > 1500$; however, the
values determined by \textsc{absori} and \textsc{pexriv} are usually
too low and poorly constrained. Furthermore, these models can only
reach a maximum $\xi$ of 3200, and so are not capable of fitting
highly ionised sources.

\item The RF fit for the high luminosity source \pks\ is not 
well constrained in $R-\xi$ space (see the lower-right panel in 
Fig.~\ref{fig:pks-results}).
This is because it is difficult to tell the difference between a highly 
reflective, highly ionised disc and a moderately ionised disc with a low 
reflection fraction. In Table~\ref{table:ross} we list fits for this
source when the reflection fraction is free and when it is fixed
at unity. Both cases give good fits to the data. We will use the 
parameters from the fit when $R$ was free in later discussions,
but the above degeneracy should be kept in mind. One possible way to 
break this degeneracy is to search for ionised edges and lines in the
soft X-ray band. These features should be present in a spectrum
from a moderately ionised disc, even with a low reflection fraction. They
would not be present at all from a highly ionised reprocessor. Obviously, the
data would need to have extremely good signal-to-noise to detect these
features and rule out alternatives such as warm absorbers. 

\item The four sources which have well constrained RF fits all have
inclination angles between 15 and 30~degrees which contradict some
earlier results (Ark~564: Turner, George \& Netzer 1999). These are
consistent with the inclination angles of broad-line
Seyfert 1 galaxies (Nandra \etal\ 1997a).  Fits to \pks\
result in large inclination angles. However, the \pks\ \fe\ line is quite 
noisy, so those inclinations are quite suspect. 

\item None of the \fe\ lines have centroid energies that are consistent
with coming from highly ionised iron. However, this does not mean that
ionised Fe is not present in the data, because the lines will be redshifted
due to relativistic blurring effects. Again, the noisiness of the \pks\ \fe\ 
line did not allow precise measurements of the line parameters.

\item Aside from \pks, all the sources are best fit with a
disc emissivity law that is proportional to $r^{-2.5}$, consistent
with the results of Nandra \etal\ (1997a). Also, most of the sources
are consistent with the emission ocurring inwards of
10~$r_g$. However, it should be emphasized that we didn't fit for
$\alpha$ and $r_{min}$, we just considered four different cases.
    
\end{enumerate}

\section{Discussion}
\label{sect:discuss}
The chief advantage of the RF model over the \textsc{pexriv} model is
that it fits the \fe\ line and the continuum simultaneously, as they
are both computed together. Since the \fe\ line is often the most
significant deviation in an otherwise featureless power-law spectrum,
this ability to fit both componants together is a significant
improvement. This is best illustrated by noting that the RF models
find a value of $\Gamma$ that is consistent with the value
computed using only high energy data, whereas models like
\textsc{pexriv} and \textsc{absori} find significantly larger values.

\begin{figure*}
\psfig{file=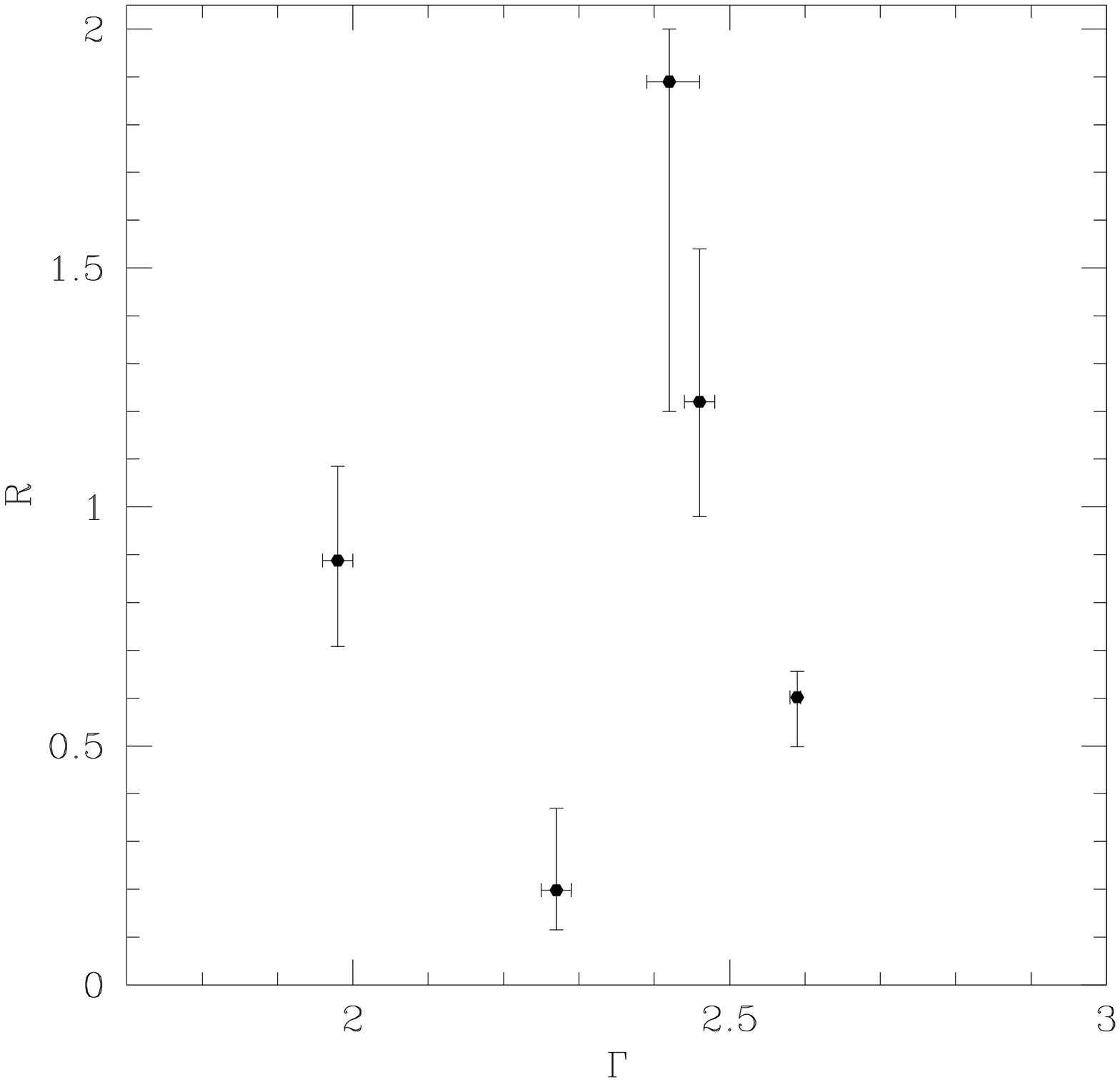,width=3.25in,silent=}
\caption{A plot of $R$ vs. $\Gamma$ using the data from 
the best fitting blurred RF model with the inclination fixed at 30~degrees. 
If there is a correlation it is very weak.}
\label{fig:r-gam}
\end{figure*}

We can now use our results to compare with ones derived from simpler models.
One of these results is an apparent correlation between the reflection
fraction $R$, and the photon index $\Gamma$, such that objects with 
larger photon indices have larger reflection fractions 
(Zdziarski, Lubi\'{n}ski \& Smith 1999). In Figure~\ref{fig:r-gam} we plot
$R$ versus $\Gamma$ for the best fitting blurred RF models (excluding \ngc,
but including the free-$R$ fit of \pks).
This figure shows that any correlation between $R$ and $\Gamma$ must be
very weak. However, it should be noted that the correlation pointed
out by Zdziarski \etal\ (1999) was mainly for sources with hard spectra;
they did not have many sources with $\Gamma > 2$.

\begin{figure*}
\psfig{file=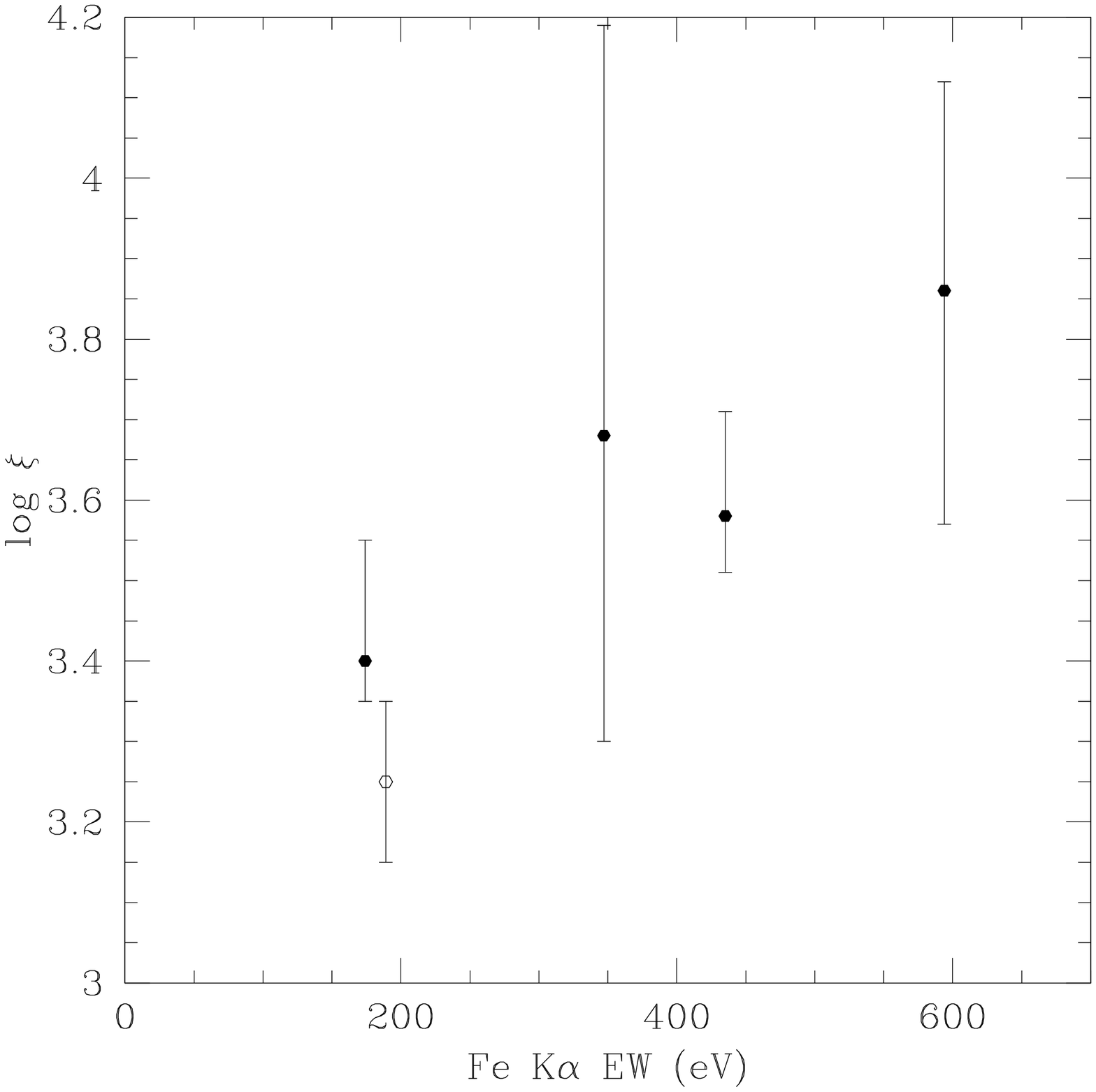,width=3.25in,silent=}
\caption{Plot of $\log \xi$ from the best fitting blurred RF models (with
the inclination fixed at 30~degrees) versus \fe\ EW, excluding \ngc. The 
source with $\Gamma < 2.0$ (\mrk) is plotted with an open symbol.}
\label{fig:ip-ew}
\end{figure*}

It was mentioned in Section~\ref{sect:model} that the constant density
RF models might not be inconsistent with the results of the variable
density models of Nayakshin \etal\ (2000) when $\Gamma > 2$. Evidence
for this interpretation is presented in Figure~\ref{fig:ip-ew}, which
plots $\log \xi$ of the best fit blurred RF models (when the
inclination was fixed at 30~degrees) of our sample versus the
equivalent width of the \fe\ line that was computed from the high
energy power-law plus gaussian fits.
This figure shows a general trend that more highly ionised sources
have larger \fe\ equivalent width. This is a typical prediction of
constant density models (e.g. Matt \etal\ 1993): as the top of the
disc becomes more ionised, emission from highly ionised Fe at 6.7 and
6.9~\kev\ will become visible and this will increase the equivalent
width of the line. Recall that most of our sources have relatively low
2--10~\kev\ luminosities (Table~\ref{table:prop}), and so we should
not observe (and, indeed, we do not) a decrease in \fe\ EW with luminosity
which is seen in higher luminosity sources (the so-called X-ray
Baldwin effect; Iwasawa \& Taniguchi 1993; Nandra \etal\ 1997b). If
our sources all had flat spectra ($\Gamma \la 2$) then the trend seen
in Figure~\ref{fig:ip-ew} would be cause for concern; however, only
one of our sources has $\Gamma < 2$ (plotted with the open symbol in
Figure~\ref{fig:ip-ew}) and this source has the smallest ionisation
parameter in our sample, consistent with the results of Nayakshin
\etal\ (2000).  The rest of the sources have $\Gamma > 2$ and have
larger ionisation parameters. Therefore, we conclude that our results
are not inconsistent with the results of Nayakshin \etal\ (2000)
despite our use of constant density models.

One other limitation of the models used in this paper is the restriction
to solar abundance. There is no reason \textit{a priori} to expect that the
abundances in accretion discs to be solar, but, then again, there is no
strong evidence that they are not solar. Changing the abundances has the
greatest effect on the \fe\ line, but, given the poor signal-to-noise in
most of the observations presented above, disentangling this effect from 
the other ones would be quite difficult. Hopefully, higher
signal-to-noise observations by \textit{XMM-Newton} and \textit{Chandra} will 
help constrain the abundances in accretion discs.  

\section{Conclusions}
\label{sect:concl}
We have fit \textit{ASCA} observations of six NLS1 galaxies with the
ionised disc models of RF, and found that five of the sources are well
fit by such a model. The only source that was not well fit by the
ionised disc model was the complicated source \ngc. The ionisation
parameter and reflection fraction of the highest luminosity source,
\pks, were not well constrained by the models. This is because the
models suffer from a degeneracy between high $R$, high $\xi$ discs and
low $R$, low $\xi$ discs. Observations with higher signal-to-noise
in the soft X-ray band might be able to break this degeneracy.

The four well constrained sources all have inclination angles typical
 of broad-line Seyfert~1 galaxies (between 15 and 30~degrees), and are
 consistent with most of the emission coming from inside 10~$r_g$. We
 find only a very weak or nonexistent correlation between the
 reflection fraction and the photon index.

While our models make the rather large simpilfying assumption that the
density of the disc is constant, because most of the sample sources
have steep spectra and will therefore show ionised features in their
reflection spectra (predicted from the non-constant density models of
Nakashin \etal\ 2000), our results are probably not unreasonable. Indeed,
the source with the flattest spectrum has the lowest ionisation
parameter in the sample and a small \fe\ EW.

These results are the strongest evidence yet that NLS1 galaxies might
posess ionised discs. This improvement comes about because the RF
models include the \fe\ line emission, can handle large ionisation
parameters, and we fit a large energy range (typically
1--10~\kev). Another ionised disc model, \textsc{pexriv}, was usually
the best out of the alternative models considered, but had difficulty
fitting the continuum and typically underestimated the ionisation
parameter. The \textsc{pexriv} model is not appropriate to fit
possibly highly ionised sources like NLS1 galaxies. These results
bolster the claim that the unique properties of NLS1 are driven by
relatively low mass black holes accreting at a high fraction of their
Eddington rate.

\section*{Acknowledgments}
We thank S.\ Vaughan for providing the \textit{RXTE} data on Ark~564,
and R.\ Ross for advice and useful discussions.  DRB acknowledges
financial support from the Commonwealth Scholarship and Fellowship
Plan and the Natural Sciences and Engineering Research Council of
Canada. KI and ACF acknowledge support from PPARC and the Royal
Society, respectively. This research has made use of data obtained
through the High Energy Astrophysics Science Archive Research Center
Online Service, provided by the NASA/Goddard Space Flight Center.


\bsp 

\label{lastpage}

\end{document}